\documentclass[journal]{IEEEtran}
\usepackage{amsmath,amsfonts}
\usepackage{algorithmic}
\usepackage{algorithm}
\usepackage{array}
\usepackage[caption=false,font=normalsize,labelfont=sf,textfont=sf]{subfig}
\usepackage{textcomp}
\usepackage{stfloats}
\usepackage{url}
\usepackage{verbatim}
\usepackage{graphicx}
\usepackage{cite}
\graphicspath{{figs/}{figures/}{pictures/}{images/}{./}}
\usepackage[dvipsnames]{xcolor}
\usepackage{graphicx}
\usepackage{tikz}
\usepackage{svg}
\usepackage{amsmath}
\usepackage[english]{babel} % To obtain English text with the blindtext package
\usepackage{enumitem}
\usepackage[breaklinks=true]{hyperref}
\newcommand{\toolName}[1]{\textit{Compendia}}
\newcommand{\q}[1]{\textit{``#1''}}
\newcommand{\eg}{e.g.,\ }
\newcommand{\req}[1]{\textit{\textbf{#1}}}
\newcommand{\ratingstats}[2]{\ensuremath{(\mu=#1)}}

\newcommand{\rvm}[1]{\textcolor{black}{#1}}

\newcommand{\hide}[1]{}

\definecolor{custombox}{HTML}{FFDAA9}

\newcommand{\outputbox}{\raisebox{0.5ex}{\fcolorbox{custombox}{custombox}{\rule{0pt}{2pt}\rule{2pt}{0pt}}}}

\newcommand{\inlineimg}[1]{\raisebox{-0.25\height}{\includegraphics[height=1.7\fontcharht\font`\A]{#1}}}

\begin{document}

\title{\toolName{}: Automated Visual Storytelling Generation from Online Article Collection}

% \author{\IEEEauthorblockN{Anonymous Authors}}


\author{Manusha Karunathilaka \orcid{0009-0001-1345-0815}, 
Litian Lei \orcid{0009-0009-5155-0313}, 
Yiming Gao \orcid{0009-0003-6090-9929}, 
Yong Wang \orcid{0000-0002-0092-0793}, 
and Jiannan Li \orcid{0000-0001-8409-4910}
        % <-this % stops a space
\thanks{M. Karunathilaka and J. Li are with Singapore Management University. Email: gmik.vidana.2023@phdcs.smu.edu.sg, jiannanli@smu.edu.sg.} 
\thanks{L. Lei, Y. Gao, and Y. Wang are with Nanyang Technological University. Email: \{litian001, gaoy0053\}@e.ntu.edu.sg, yong-wang@ntu.edu.sg.}
\thanks{J. Li and Y. Wang are the corresponding authors.}
}

\markboth{Journal of \LaTeX\ Class Files,~Vol.~14, No.~8, August~2021}%
{Shell \MakeLowercase{\textit{et al.}}: A Sample Article Using IEEEtran.cls for IEEE Journals}


\maketitle

\begin{abstract}
In the digital age, readers value quantitative journalism that is clear, concise, analytical, and human‑centred. 
To understand complex topics, they often piece together scattered facts from multiple articles. 
Visual storytelling can transform fragmented information into clear, engaging narratives, yet its use with unstructured online articles remains largely unexplored.
To fill this gap, we present \toolName{}, an automated system that analyzes online articles in response to a user’s query and generates a coherent data story tailored to the user’s informational needs. 
\toolName{} addresses key challenges of storytelling from unstructured text through two modules covering: \textit{Online Article Retrieval}, which gathers relevant articles; \textit{Data Fact Extraction}, which identifies, validates, and refines quantitative facts; \textit{Fact Organization}, which clusters and merges related facts into coherent thematic groups; and \textit{Visual Storytelling},
which transforms the organized facts into narratives with visualizations in an interactive scrollytelling interface.
We evaluated \toolName{} through a quantitative analysis, confirming the accuracy in fact extraction and organization, and through two user studies with 16 participants, demonstrating its usability, effectiveness, and ability to produce engaging visual stories for open-ended queries.
\end{abstract}

\begin{IEEEkeywords}
Data Storytelling, Scrollytelling, Text/Document Data
\end{IEEEkeywords}

\section{Introduction}

In today’s information-rich era, people rely on diverse online sources, such as news articles, blogs, reports, and expert analyses, to stay updated on events and facts~\cite{case2016looking, purcell2010understanding}.
Audience studies indicate that readers particularly value quantitative journalism that is constructive, concise, analytical, and human‑centred~\cite{stalph2024exploring},
and visualizations have been widely used to convey such quantitative information clearly and engagingly~\cite{yu2024morethandatastory}.
Following this trend and growing \rvm{audience demand from news consumers and analysts alike}, 
major information outlets, such as The New York Times, The Guardian, and Pew Research Center\footnote{\url{https://www.pewresearch.org/}},
% ~\cite{pewresearchResearchCenter},
have made statistics and explanatory visual journalism a core part of their reporting.
Some of them maintain dedicated data‑driven sections, 
including The Guardian’s Datablog\footnote{\url{https://www.theguardian.com/news/datablog}}
% ~\cite{theguardianDatablogGuardian}
% Financial Times’ Datawatch~\cite{ftDatawatch}, 
and New York Times’ Upshot\footnote{\url{https://www.nytimes.com/international/section/upshot}}.

\rvm{However, when approaching complex phenomena with multiple facets, readers, such as news consumers and analysts, often have to consult multiple sources to gain a more complete picture~\cite{purcell2010understanding}.
For example, a reader who is trying to understand the climate impact of artificial intelligence (AI) might need to integrate technology landscapes, energy estimates, and carbon footprint assessments,
which are scattered across news articles, company white papers, and government reports. 
Similarly, an analyst studying global poverty might draw on income distributions and social welfare policies from multiple government websites and think tank reports.   
In practice, readers must iteratively refine web search queries, carefully scan extensive contents to extract relevant quantitative information, and then relate findings across sources to construct a coherent understanding~\cite{marchionini2006exploratory}.
This manual effort creates substantial cognitive burden~\cite{stalph2024exploring}, often obscuring overarching themes and making it difficult to reconcile related facts into a clear mental model.}

To ease the burden, existing aggregator systems, such as Google News, Newsblaster\cite{mckeown2002tracking}, and NewsBird~\cite{hamborg2017matrix}, consolidate content from multiple articles into a single interface. 
Yet, they tend to focus on analyzing surface‑level information, such as
headlines and short summaries, 
\rvm{leaving readers to perform the cognitively demanding task of synthesizing quantitative information scattered across sources.}
Similarly, AI‑powered search engines like Perplexity\footnote{\url{https://www.perplexity.ai/}}
produce cited summaries of web content, but they remain largely text‑centric and lack the quantitative synthesis and visual scaffolding \rvm{that readers strongly prefer when navigating complex, data-rich topics~\cite{stalph2024exploring}.}

Visual storytelling transforms complex topics into coherent narratives that combine quantitative evidence with visual presentations, enhancing the comprehension, engagement, and recall of the underlying data and insights~\cite{borkin2015beyond, dove2012narrative, segel2010narrative}.
\rvm{Automated visual story generation saves significant manual efforts, but prior work on automated visual storytelling required structured data (\eg tables)~\cite{wang2019datashot, shi2020calliope, islam2024datanarrative} or a single article as source~\cite{hullman2013contextifier, leake2020generating} and lacked the ability to integrate information across multiple articles~\cite{yu2024morethandatastory}.
We address this gap by combining automated text analysis with visual data story generation over online article collections, aiming to help readers synthesize quantitative facts and form a coherent mental model of a topic in support of their analytical tasks.}

\begin{figure*}[!htbp] % Use figure* for a two-column figure, [t] for top of the page
  \centering % More compact than using \begin{center} ... \end{center}
   \setlength{\abovecaptionskip}{0.1em} 
  \includegraphics[width=\textwidth]{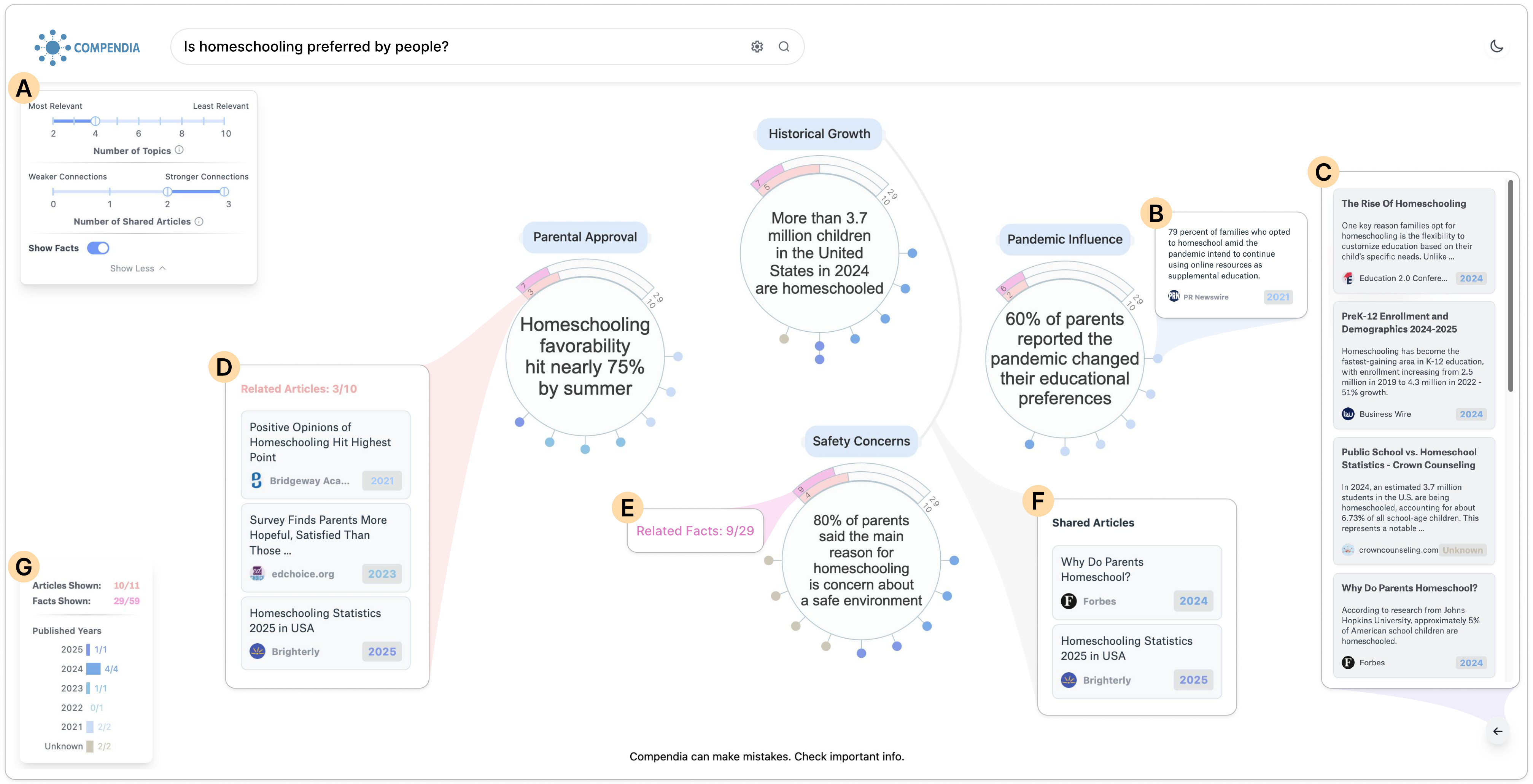} % Use \textwidth for spanning across two columns
   \caption{%
    \toolName{} transforms the query \q{Is homeschooling preferred by people?} into a structured data story by extracting, clustering, and visualizing key facts using unstructured data from a collection of online articles.
 \textbf{Thematic Overview} uses \textbf{Thematic Circles} to visualize clustered facts across different themes. (A) Filter widget provides control over the overview, (B) Detailed fact panel provides fact content and source details, (C) Articles panel presents all retrieved articles relevant to the story, (D) Related Articles panel displays articles relevant to the topic, (E) Related facts panel provides the number of facts belonging to the topic, (F) Shared articles panel lists sources covering multiple aspects of the topic, and (G) Summary panel displays article and fact statistics.}
  \vspace{-1.5em} 
  \label{fig:teaser}
\end{figure*}

Generating visual data stories from a collection of online articles remains largely unexplored and intrinsically challenging.
The challenges 
of generating a coherent data story from 
online article collections 
originate from the fact that these data sources are often vastly \textbf{diverse} and \textbf{unstructured}.
First, relevant data facts are often scattered across multiple articles and sections, mixed with irrelevant content, and expressed in varied formats and units (\eg 3.7K/3700/3,700).
\textit{Extracting} data facts from these sources requires a systematic approach to identify key insights while effectively filtering out irrelevant or noisy information.
Second, online articles often differ significantly in their themes, angles, and writing styles.
Such diversity enhances the richness of data stories but also complicates the task of \textit{organizing} fragmented facts into clear, coherent narratives.
Third, the large quantity and structural complexity of the extracted facts require a narrative strategy and a visual \textit{presentation} design that can preserve the richness of the information without overwhelming readers.

To tackle these challenges, we present \textbf{\toolName{}}, an automated system that retrieves and analyzes collections of online articles in response to a user query and generates a coherent, meaningful visual data story.
We propose a framework that includes two main modules: the \textbf{Data Fact Extraction and Organization} module and the \textbf{Visual Storytelling} module, leveraging Large Language Models (LLMs) to overcome the complexities of unstructured online text and convert it into a coherent data story. 
The \textit{Data Fact Extraction and Organization} module consists of:
\textbf{(1) \textit{Online Article Retrieval}} gathers relevant articles in bulk using web scraping; 
\textbf{(2) \textit{Data Fact Extraction}} filters out irrelevant text and extracts query‑relevant quantitative facts, capturing not only their values but also associated units and contextual information to preserve meaning across diverse formats; and
\textbf{(3) \textit{Fact Organization}} clusters multifaceted data facts into thematic groups, distinguishes core from supporting facts, and enhances clarity by merging closely related facts, resolving ambiguities, and ensuring each group forms a coherent narrative unit.
The \textit{Visual Storytelling} module weaves narrative units into a coherent story, presenting a \textbf{Thematic Overview} 
\rvm{allowing users to form a high-level mental model of the topic,} followed by interactive \textbf{Story Views} in an \q{overview first, details on demand} manner~\cite{shneiderman2003eyes}.
Users then drill down into the story through intuitive page scrolling, leveraging the scrollytelling technique~\cite{tjarnhage2023impact,seyser2018scrollytelling}.

We evaluated \toolName{} through 
(1) quantitative evaluation, confirming its ability to accurately identify, extract, and organize information from diverse article collections, 
and 
\rvm{(2) two qualitative user studies with 16 participants:
an initial perception study establishing \toolName{}'s effectiveness and usability, 
and an open-ended query exploration study demonstrating its ecological validity and utility in real-world use.}

The major contributions of this paper can be summarized as follows:
\begin{itemize}
    \item \textbf{An end-to-end framework} that automatically generates relevant, accurate, and well-organized data stories by extracting and structuring highly varied unstructured textual data from online articles.
    \item \textbf{An interactive visualization system}, \toolName{}, demonstrating the feasibility of generating and presenting data stories from online articles based on user queries.
    \item \textbf{A comprehensive evaluation}, including a quantitative evaluation and two user studies, confirming \toolName{}'s information accuracy and its effectiveness in producing coherent, informative, and engaging data stories.
\end{itemize}

\section{Related Work}
The related work falls into two categories: data storytelling, and information retrieval and extraction techniques.

\vspace{-1em}
\subsection{Data Storytelling}
A visual data story is a series of story elements arranged in a meaningful sequence, forming a narrative visualization~\cite{segel2010narrative, lee2015more}. 
Research in this area has examined diverse genre from annotated charts and comics to timelines and animated narratives, and has shown that storytelling enhances memory, engagement, comprehension, and communication~\cite{borkin2015beyond, dove2012narrative, segel2010narrative, shao2024data}.
Furthermore, data stories have been shown to enhance both the efficiency and effectiveness of comprehension tasks, outperforming conventional visualizations~\cite{shao2024data}.
Recent studies further refine storytelling by exploring narrative roles, AI support levels, and human-AI collaboration~\cite{tong2018storytelling, chen2023does, li2024we, he2024leveraging}.
Yet, creating those data stories poses significant challenges. 

Early tools let users manually build these visual stories using templates, such as annotated charts~\cite{ren2017chartaccent},
% ~\cite{tableauTableauBusiness, ren2017chartaccent}, 
infographics~\cite{cui2021mixed}, 
% ~\cite{wang2018infonice, cui2021mixed}, 
timelines~\cite{offenwanger2023timesplines}, 
% ~\cite{brehmer2019timeline, offenwanger2023timesplines}, 
and data comics~\cite{kang2021toonnote}. 
% ~\cite{kim2019datatoon, kang2021toonnote}. 
Later, systems like Idyll~\cite{conlen2018idyll} and VisFlow~\cite{sultanum2021leveraging} supported dynamic formats such as scrollytelling and slideshows, but they often required design and coding skills.
More recent semi-automated tools add smart features that help link text and visuals~\cite{latif2021kori}, infographic recommendation~\cite{tyagi2021user}, slide generation~\cite{li2023notable}, and data story editing~\cite{sun2022erato}, but they are primarily designed for expert users like data analysts and designers.

Fully automated systems like DataShot~\cite{wang2019datashot} and Calliope~\cite{shi2020calliope} can build stories from structured data (e.g., tables and spreadsheets), but they often miss deeper meanings. 
Similarly, recent research has extended this automation towards generating scrollytelling narratives from structured data~\cite{lu2021automatic}. 
New approaches that use LLMs for narrative generation~\cite{sultanum2023datatales, cheng2024snil, islam2024datanarrative} still rely mostly on structured data, leaving the rich information embedded in unstructured text (e.g., text articles) data underexplored. 
Furthermore, most of these tools do not explicitly support the direct communication of data stories to audiences and often lack mechanisms to incorporate user intent, limiting their effectiveness in tailoring the narrative to the user's needs~\cite{li2024we}.

Our work addresses these challenges by automatically generating data stories from unstructured text sources such as online articles, tailored to users' informational needs.

\vspace{-0.5em}
\subsection{Information Retrieval and Extraction}
Information Retrieval (IR) focuses on retrieving relevant information from large-scale text corpora in response to user queries.
Traditional IR systems rely on statistical and probabilistic models, such as Term Frequency--Inverse Document Frequency (TF--IDF)~\cite{salton1988term} and the Okapi BM25 ranking function~\cite{robertson2009probabilistic}.
These methods compute lexical similarity between queries and documents, providing a foundation for early search engines and document retrieval systems. 
However, these approaches struggle with semantic understanding, as they primarily depend on keyword matching rather than contextual comprehension.
More recent neural retrieval methods, particularly those based on transformer architectures like BERT~\cite{devlin2019bert}, have improved search effectiveness by leveraging contextual embeddings. However, these methods still require extensive domain-specific training and fine-tuning, making them less adaptable to diverse and evolving information needs.

Beyond retrieval, Information Extraction (IE) is essential in structuring retrieved content by identifying key entities, relationships, and events.
Traditional IE methods typically rely on separate models for different tasks, such as Named Entity Recognition (NER), Relation Extraction (RE), and Event Extraction (EE)~\cite{nadeau2007survey}. 
However, the need for multiple independent models for different tasks made these approaches resource-intensive. 
Recent advancements in LLMs, such as GPT-4~\cite{achiam2023gpt}, have introduced generative IE methods that offer a more scalable alternative by generating structured knowledge directly from text, reducing the need for task-specific models~\cite{li2024matching}.
This shift allows more flexible and efficient extraction from complex, evolving data.

In the context of LLMs, prompt engineering is fundamental for guiding the behavior of LLMs to optimize their performance~\cite{liu2023pre}. 
Techniques such as Chain-of-thought (CoT) prompting~\cite{wei2022chain}, which provides a structured reasoning process to direct model responses, and Few-shot prompting~\cite{brown2020language}, which utilizes in-context learning by incorporating example demonstrations, are commonly applied to improve the performance. 
\toolName{} leverages LLMs to enhance information retrieval and extraction. 
Carefully designed prompts ensure that retrieved content is both relevant and structured to support automated data story generation.
\section{Compendia Framework Overview}
\rvm{\toolName{} is the first work to achieve \textit{automated} visual storytelling generation from online articles, to the best of our knowledge. 
Yet, there has been much research on visual storytelling~\cite{dork2012pivotpaths,segel2010narrative,tjarnhage2023impact, seyser2018scrollytelling}, exploratory research~\cite{marchionini2006exploratory, white2009exploratory, hornbaek2011notion, kules2008users}, and information visualization principles~\cite{shneiderman2003eyes}. 
To guide our framework and visualization designs, we have established the following visualization design requirements by conducting an extensive literature review:
\begin{description}
  \item \textbf{(R1) Provide an overview of topic clusters:}
  During the exploratory research or query, users need a quick overview of the topics covered in the online articles to orient themselves and guide further inquiry~\cite{marchionini2006exploratory, white2009exploratory}. This is well-aligned with the visualization mantra \q{overview first, details-on-demand}~\cite{shneiderman2003eyes}. 
  Prior work shows that thematic overviews reduce cognitive load and improve sensemaking~\cite{hornbaek2011notion, kules2008users}, especially when information is grouped by topic rather than listed flatly~\cite{hearst1996reexamining, zamir1999grouper}.  
  \item \textbf{(R2) Reveal the connections among topics:}  
  Revealing the correlation between the topics covered help users gain a more structured understanding when exploring multiple related documents~\cite{dork2012pivotpaths}. 
  Prior work has also shown that linking articles by shared topics helps trace storylines across documents~\cite{das2005connecting, sun2007topic}. 
  \item \textbf{(R3) Enable smooth navigation between topics and data facts:}
  There are often multiple topics and data facts covered in a large collection of online articles. 
  Existing research on visual storytelling~\cite{segel2010narrative,tjarnhage2023impact, seyser2018scrollytelling} has emphasized the importance of maintaining data fact continuity and smooth navigation to reduce cognitive load. 
  Also, many online data storytelling examples~\cite{nytimesGreatFlood, puddingWhatCity} have often used a map metaphor to enable the transition and navigation among different topics. 
  \item \textbf{(R4) Support visualizations with narratives: }
  Data visualizations coupled with narrative text help readers interpret quantitative facts more effectively than raw facts or isolated charts~\cite{segel2010narrative, hullman2011visualization}. 
  Furthermore, weaving facts into story-driven visuals enhances comprehension, retention, and engagement~\cite{shao2024data}.
  \item \textbf{(R5) Enable direct access to facts and their sources:}  
  To promote transparency, users must be able to trace each fact to its original source article.  
  Prior work shows that source attribution improves credibility and trust~\cite{solovev2025references, diel2021news, borah2014hyperlinked}, with interfaces like TileBars enabling direct access to source passages~\cite{hearst1995tilebars}.  
  We aim to ensure that every fact is displayed with its original details, citation, and publication year.
\end{description}}

\rvm{To satisfy these design requirements, the \toolName{} framework consists of two core modules:} the \textit{Data Fact Extraction and Organization} module and the \textit{Visual Storytelling} module, as shown in Fig.~\ref{fig:framework}.
Unlike the prior systems such as DataShot~\cite{wang2019datashot}, which operate on structured table data, \toolName{} works directly on natural language text where quantitative numbers and related descriptions are often scattered across multiple documents. 
This introduces challenges such as noisy paragraphs, vague expressions (\eg \q{this year,} \q{currently}), fragmented statements, and inconsistencies in units or time frames. 
To address these issues, the two modules progressively extract, validate, and organize the data facts, finally presenting them as a data story.

The first module (Fig.~\ref{fig:framework}\inlineimg{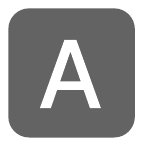}) retrieves and processes unstructured text from online articles through three stages.
The \textbf{Online Article Retrieval} 
expands the user query with variations and automatically searches and scrapes relevant articles.
The \textbf{Data Fact Extraction} then filters paragraphs, identifies and extracts quantitative facts with their values, units, and context.
Finally, \textbf{Fact Organization} clusters semantically related facts and merges coherent facts.
These stages parallel the workflow of DataShot~\cite{wang2019datashot}, but extend it from structured tables to unstructured articles: replacing direct table parsing with LLM-guided fact extraction, redefining fact composition through clustering and merging.

The second module (Fig.~\ref{fig:framework}\inlineimg{B.pdf}) transforms structured fact clusters into interactive narratives, weaving them into a coherent story flow that combines visualization with scrollytelling for effective presentation.
This module consists of two views, beginning with the  \textbf{Thematic Overview} (in Fig.\ref{fig:framework}\inlineimg{B.pdf}), which provides a high-level summary of all extracted data facts as clusters; 
and followed by the \textbf{Story View} (in Fig.\ref{fig:framework}\inlineimg{B.pdf}), accessible by scrolling or selecting a cluster, which enables detailed exploration of data facts in that cluster.
We discuss the two modules in the following sections.

\section{Data fact extraction and organization}\label{sec:framework}
\begin{figure*}[!htbp] % Use figure* for a two-column figure, [t] for top of the page
  \centering % More compact than using \begin{center} ... \end{center}
   \setlength{\abovecaptionskip}{0.1em} 
  \includegraphics[width=\textwidth]{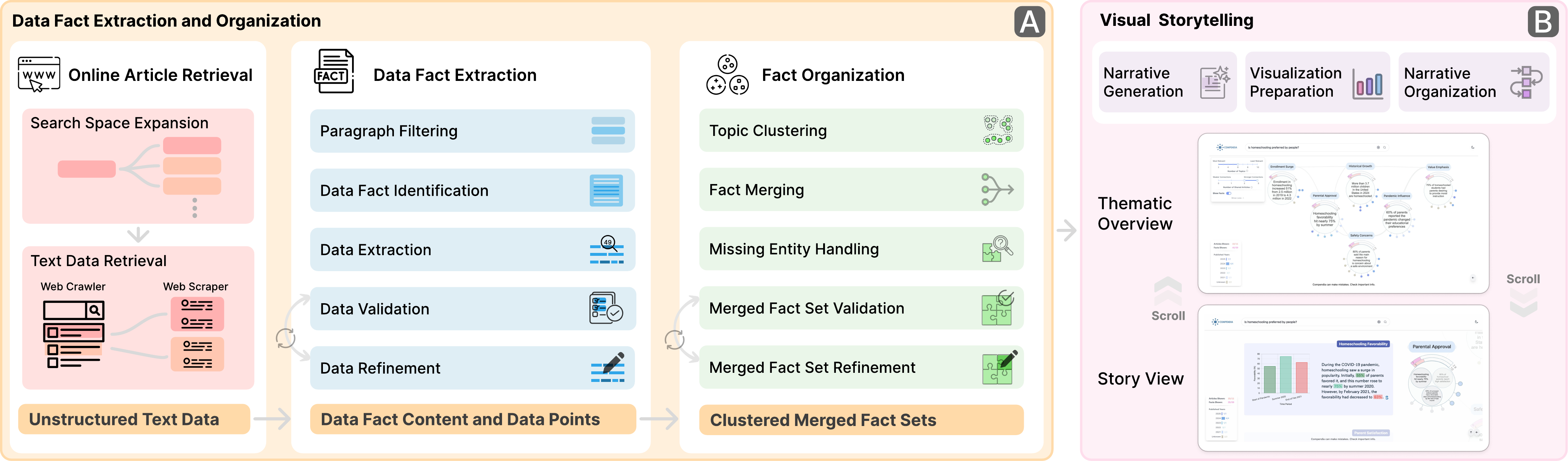} % Use \textwidth for spanning across two columns
   \caption{%
  The framework of our system, \toolName{}. It consists of two main modules: A) the Data Fact Extraction and Organization module, which includes Online Article Retrieval, Data Fact Extraction, and Fact Organization; and B) the Visual Storytelling module, which prepares the data for display and presents it as an interactive scrollytelling story.
  Each stage involves steps performed in the given order, with iterative validation and refinement phases.
  The \protect\outputbox{} boxes highlight the outputs at each stage.
  }
  \vspace{-1.5em} 
  \label{fig:framework}
\end{figure*}

Our data fact extraction and organization module 
(Fig.~\ref{fig:framework}\inlineimg{A.pdf}) processes unstructured text from online articles and transforms them into structured quantitative facts that form the foundation for coherent storytelling.
In this section, we describe each stage of the module and
the implementation details.

\vspace{-1em}
\subsection{Online Article Retrieval}
The online article retrieval stage collects the information sources for data stories.
It begins with a user-provided natural language query, which reflects their informational intent. 
As search results are increasingly affected by spam and ranking manipulation~\cite{bevendorff:2024a}, 
we expand the original query with LLM‑generated variants to improve coverage and reduce unrelated content~\cite{bevendorff:2024a}.
Based on these queries, \toolName{} uses an online search engine to retrieve relevant articles, emulating how a user would search for information.
\rvm{The source article is tracked at every stage for transparency (\req{R5}).}

\vspace{-1em}
\subsection{Data Fact Extraction}
\toolName{}'s focus is to find and extract quantitative evidence that supports the user's query. 
This stage aims to extract  relevant data facts from the retrieved article text, through multiple phases to progressively filter, identify, validate, and refine the data, ultimately generating a well-structured dataset, \rvm{which is easier to visualize (\req{R4})}.

\subsubsection{Paragraph Filtering}
This phase filters the paragraphs related to the user's query from scraped article text, where paragraphs preserve more context than
individual sentences and enable better relevance assessment and
disambiguation~\cite{barzilay2004catching}. 
These texts often contain extraneous information such as headers, footers, references, advertisements, and unrelated content. 
The system first prompts an LLM to filter the paragraphs using the article title to remove extraneous information,
and then re-filters the paragraphs based on the relevance to the query to align with the user’s search intent, producing a set of \textbf{filtered paragraphs} as foundation
for subsequent phases.

\subsubsection{Data Fact Identification} \label{subsubsec:factidentification}
This phase focuses on extracting relevant quantitative information, termed \textbf{fact content}, from the \textit{filtered paragraphs} with the aid of an LLM.
This process introduces several key challenges.
First, distinguishing data facts from factual statements is challenging, 
as not all factual sentences contain analytically useful quantitative information. 
For example, the sentence \q{Joshua and Samuel are two popular names for boys in 2004} represents a categorization-type data fact~\cite{wang2019datashot}, 
but lacks a numerical value for quantitative analysis or visualization.
Therefore, we primarily focus on identifying and isolating quantifiable facts, such as numerical data, measurements, and statistics, while excluding subjective opinions and interpretations.
Second, multiple facts may appear within a single paragraph, often interwoven with contextual or explanatory content.
Specifically, we focus on identifying sentence or phrase level data facts, where each fact contains at least one data value. 
For example, \q{More than 3.7 million children in U.S. in 2024 are homeschooled} illustrates a sentence-level data fact with a clearly stated numerical value.
Identifying such sentence or phrase level fact content is also crucial for downstream analysis, particularly for grouping similar facts, as outlined in Section~\ref{sec:factmerging}.
To systematically identify data facts, we embed Datashot’s~\cite{wang2019datashot} quantifiable fact type definitions and examples into the prompt.

\subsubsection{Data Extraction}\label{subsubsec:dataextraction}
This phase extracts numerical data by parsing each \textit{fact content}, termed 
\textbf{data points} using an LLM.
Several challenges arise during this process.
First, numerical values appear in varied writing styles, such as \q{3,700,000}, \q{3.7 million}, or \q{3.7M}. 
Second, a single fact may contain multiple numerical values, each requiring separate extraction and contextual alignment.
For example, the sentence \q{From 1999 to 2020, the number of homeschooled students in the U.S. increased from 850,000 students to 2.5 million students} 
includes two distinct values associated with different timestamps, which must be captured individually while preserving their temporal semantics.
In response, we treat each numerical value individually as a \textbf{data point}, extracting them into a unified structured representation consisting of three key fields: \textbf{label}, \textbf{value}, and \textbf{unit}.
This returns a set of data points per fact content. 
For example, given a \textit{fact content} \q{The number of homeschoolers in the U.S. is 3.7 million}, the system extracts the label \q{Homeschooled Children}, value \q{3.7} with the unit \q{million.}
Finally, contextual ambiguity in the text can make accurate data extraction challenging.
For example, \q{this year} could be misinterpreted as the current year (2025) when the article was actually published in 2024. 
To resolve such ambiguities, we incorporate article metadata alongside the content in the prompt, which helps clarify ambiguous data.

\subsubsection{Data Validation}
This phase assesses the correctness, consistency, and completeness of extracted \textit{fact contents} (Section \ref{subsubsec:factidentification}) and \textit{data points} (Section \ref{subsubsec:dataextraction}) using an LLM and suggests corrections for discrepancies and errors.
The process begins by prompting an LLM to verify  \textit{fact contents} against the original text and identify discrepancies.
Next, it validates the \textit{data points}, prompting an LLM to check their correctness and completeness against the original text data and ensure the unit consistency within data points of a fact.
For example, if a paragraph states, \q{The number of homeschoolers in the U.S. is 3.7 million,} but the extracted data point's value is 3, the system detects the error and suggests correcting the value to 3.7. 
Similarly, if units such as \q{billion} and \q{million} 
appear inconsistently within a fact’s data points,  
the system flags the inconsistency and recommends consistently normalizing them.

\subsubsection{Data Refinement}
This phase corrects errors in \textit{fact contents} and \textit{data points} by prompting an LLM based on validation feedback.
The system reviews the feedback by comparing each fact with the source text and then applies minimal edits to preserve original context. 
Common fixes include adjusting numerical values, normalizing units, and resolving misinterpretations. 
The validation and refinement phases are performed iteratively to ensure the final structured fact data is correct, consistent, and complete.

\vspace{-1em}
\subsection{Fact Organization}

The facts from diverse online articles are often fragmented and inconsistent in style, and vary in detail and focus. 
This stage clusters facts into themes \rvm{(\req{R1})} and merges related facts into coherent narratives  \rvm{(\req{R4})}, reducing redundancy while preserving diversity, yielding a consistent and comprehensive view of the themes drawn from multiple sources.

\subsubsection{Thematic Clustering}\label{sec:topicclustering}

This phase facilitates clustering the facts into themes, returning \textbf{clustered facts} \rvm{(\req{R1})}.
Those extracted facts originate from a wide range of online articles, each written with different styles, emphases, and levels of detail. 
As a result, the collected facts are  inconsistent in granularity, with some capturing high-level summaries, while others offering fine-grained statistics, yet often sharing common themes.
Identifying such thematic overlap across facts is important to organize the fragmented facts.

To organize the extracted facts into meaningful groups, we perform thematic clustering.
The aggregate length of the text facts often exceeds the context window limits of LLMs, and chunking the input to fit these limits tends to degrade clustering quality~\cite{liu2024lost}.
To better support large-scale clustering in an unsupervised setting, we adopt an embedding-based approach using the Gaussian Mixture Model (GMM)~\cite{katz2024knowledge}.
Specifically, we first obtain vector representations of the \textit{fact contents} using an LLM-based embedding service, and then apply GMM to cluster these embeddings based on their semantic similarity, 
producing clustered facts.
We then generate a concise \textbf{cluster topic} using an LLM for each cluster to help users quickly interpret and compare them.
However, generating topics independently for clusters can result in ambiguity or overlap due to their conciseness.
To improve clarity and differentiation across topics, we first prompt an LLM to generate a brief summary for each cluster.
We then provide these summaries, together with the initial topics, back to the LLM to refine them into a final set of distinct, thematically coherent topics.

Furthermore, we calculate each fact’s relevance score as the cosine similarity between its embedding and the query embedding. 
This relevance score helps to identify the most representative fact (highest relevance score) and the top three facts in a cluster. 
Moreover, we calculate each cluster’s relevance score as the average of the fact relevance scores within the cluster, which is then used for filtering (Section~\ref{sec:fliterwidgets}).

\subsubsection{Fact Merging}\label{sec:factmerging}
This phase consists of two steps: first, resolve redundancies by constructing sets of factually similar facts as \textbf{fact sets}; and second, merging semantically similar fact sets into \textbf{merged fact sets}, within a cluster. 

\textbf{Resolve Redundancy.}
Multiple articles often express the same fact in different wording or styles, yet convey equivalent underlying factual content.
For example, \q{3.7 million children were homeschooled in 2024} and \q{In 2024,  3,700,000 students homeschooled.}
To handle the redundancy, we use an LLM to construct \textbf{fact sets} within each cluster based on their factual similarity in \textit{fact content} and \textit{data points}.
We observe that semantically similar statements may include conflicting values. Resolving these contradictions requires more sophisticated fact checking, with potential approaches noted in Section~\ref{sec:limitations}.

\textbf{Merge Fact Sets.}
The extracted facts are often fine-grained, captured at the sentence or phrase level.
While this granularity is beneficial for identifying similar facts as discussed above, it also introduces significant challenges such as fragmented information that disrupts narrative coherence and leads to cognitive overload, when facts are presented in isolation. 
In response, we propose a merging process that combines semantically similar \textit{fact sets} within each cluster. 
These \textbf{merged fact sets} form the basis of the \textit{Narrative Units} discussed in the Section~\ref{sec:narrativeunit}.
This improves the readability and reduces the number of narratives that users have to scroll through.

This merging is challenging due to variations in perspectives, time frames, and units, while also needing to ensure the compatibility with the visualizations that support the narrative \rvm{(\req{R4})}.
To achieve coherence and compatibility, our merging process follows a structured approach within a prompt, incorporating \textbf{Unit Consistency}, where all facts use the same measurement units; \textbf{Thematic Relevance}, where facts contribute to a cohesive story; \textbf{Temporal Consistency}, where facts share logically connected time frames; and \textbf{Visualization Readiness}, where data share a consistent x- and y-axis.

\underline{\textit{Unit Consistency.}}
The fact sets are grouped by unit type to maintain consistency, while normalizing the units when necessary (\eg \q{3,000,000} → \q{3 million}). 
The fact sets with different measurements or unit types are not merged.
For example, \q{Three million children are homeschooled} cannot be merged with \q{20\% of students were homeschooled,} as one uses a count and the other a proportion.

% \vspace{-0.5em}
\underline{\textit{Thematic Relevance.}}
The \textit{fact sets} are merged when they share a common theme.
For example, \q{23.1\% said that the reason for homeschooling was the child’s special needs} and \q{15.6\% of parents said that the child had a physical or mental problem} can be merged as: \q{23.1\% of parents cited their child’s special needs as the reason for homeschooling, while 15.6\% pointed to physical or mental health challenges.}

% \vspace{-0.5em}
\underline{\textit{Temporal Consistency.}}
The \textit{fact sets} are merged only when they share compatible time ranges; those from different periods remain separate.
For example, \q{Homeschooling increased by 25\% between 2020 and 2021} cannot be merged with \q{Homeschooling rates have doubled over the past decade,} 
as the first reflects a short‑term trend and the second a long‑term pattern.

% \vspace{-0.5em}
\underline{\textit{Visualization Readiness.}}
Finally, we assess whether the \textit{merged fact set} is compatible with a visual representation, ensuring that the data can be aligned along common axes (\eg consistent time or category labels on the x-axis and comparable values on the y-axis).

% \vspace{-1em}
\subsubsection{Missing Entity Handling}\label{secs:missingentity}
After obtaining the \textit{merged fact sets}, 
we often encounter missing entities, such as geographical locations, time references, or subject identifiers, that are essential for clarity. 
These gaps can make the narrative ambiguous or incomplete and typically arise from the fragmented nature of fact extraction, 
where individual sentences may omit context present in the original source.
Identifying all necessary contextual information during the initial extraction stage is often infeasible due to the complexity and variability of how information is presented across different articles.
For example, consider the following two facts within a merged group: \q{23.1\% of parents in the U.S. cited special needs as a reason for homeschooling} and \q{15.6\% of parents said that the child had a physical or mental problem.} While the first fact explicitly mentions the U.S., the second fact lacks a geographical reference, making it ambiguous.

To tackle this, we employ a multi-step approach to identify and fill missing entities. 
First, we use Named Entity Recognition (NER) to detect and classify entity types in each fact.
Next, we compare the entity types across facts within each \textit{merged fact set} to detect missing entities.
If an entity is present in one fact but absent in others, our algorithm flags it as a potential missing entity. 
To fill those missing entities, we leverage an LLM
to analyze the original article content and infer the missing entities.
Finally, these \textit{merged fact sets} undergo an iterative validation and refinement phase similar to the previous stage.
The validation checks their correctness against the original article and provides feedback, while the refinement applies corresponding edits.

\subsection{Implementation Details}
We employ Google Search for article retrieval, 
using SearchAPI\footnote{\url{https://www.searchapi.io}} to collect organic Search Engine Results Page (SERP) and Serper API\footnote{\url{https://serper.dev/}} to extract page text, 
in compliance with source site terms and copyright laws.
For the LLM mentioned in both Sections~\ref{sec:framework} and~\ref{sec:visualization}, we use OpenAI’s \textit{gpt-4o-2024-08-06} model, 
which supports structured outputs\footnote{\url{https://openai.com/index/introducing-structured-outputs-in-the-api/}} for consistent formatting. 
We define the output schema for each LLM task and apply techniques such as Chain-of-Thought (CoT), prompt chaining, and few-shot prompting. 
Detailed prompts are provided in the supplemental materials.
For the clustering, 
we utilize OpenAI’s \textit{text-embedding-3-large} model for vector embeddings and the Gaussian Mixture Model (GMM) method, 
which supports soft clustering in an unsupervised manner~\cite{katz2024knowledge}. 
To accommodate screen size constraints and mitigate information overload when visualizing clusters as discussed in Section~\ref{secs:thematicoverview}, 
we limit the maximum number of clusters to 10. 
For the NER task (Section~\ref{secs:missingentity}), we use SpaCy’s transformer-based model, \textit{en\_core\_web\_trf}\footnote{\url{https://spacy.io/models/en\#en_core_web_trf}}, due to its robust performance and comprehensive entity coverage.
We have released  \toolName{}’s source code here: \url{https://github.com/compendia-project} for public use.

\section{Visual Storytelling}\label{sec:visualization}

Our visual storytelling module builds upon the above module and consists of a \textbf{Thematic Overview} and \textbf{Story View}, as shown in Fig.~\ref{fig:framework}\inlineimg{B.pdf}. 
The \textit{Thematic Overview} provides users with a brief visual summary of major topics related to their query, and the \textit{Story View} enables users to explore detailed data facts in an interactive and engaging manner.

\vspace{-1em}

\subsection{Thematic Overview}\label{secs:thematicoverview}
The \textbf{Thematic Overview} (Fig.~\ref{fig:teaser}) groups extracted facts into semantically-coherent topics, allowing users to gain a quick summary of all the relevant topics for a given query (\req{R1}).
To avoid overwhelming users, the \textit{Thematic Overview} integrates a \textbf{Filter Widget} that allows the users to simplify the view and focus on what matters the most.

\begin{figure}[!t]
\centering
   \setlength{\abovecaptionskip}{0.1em} 
\includegraphics[width=\columnwidth]{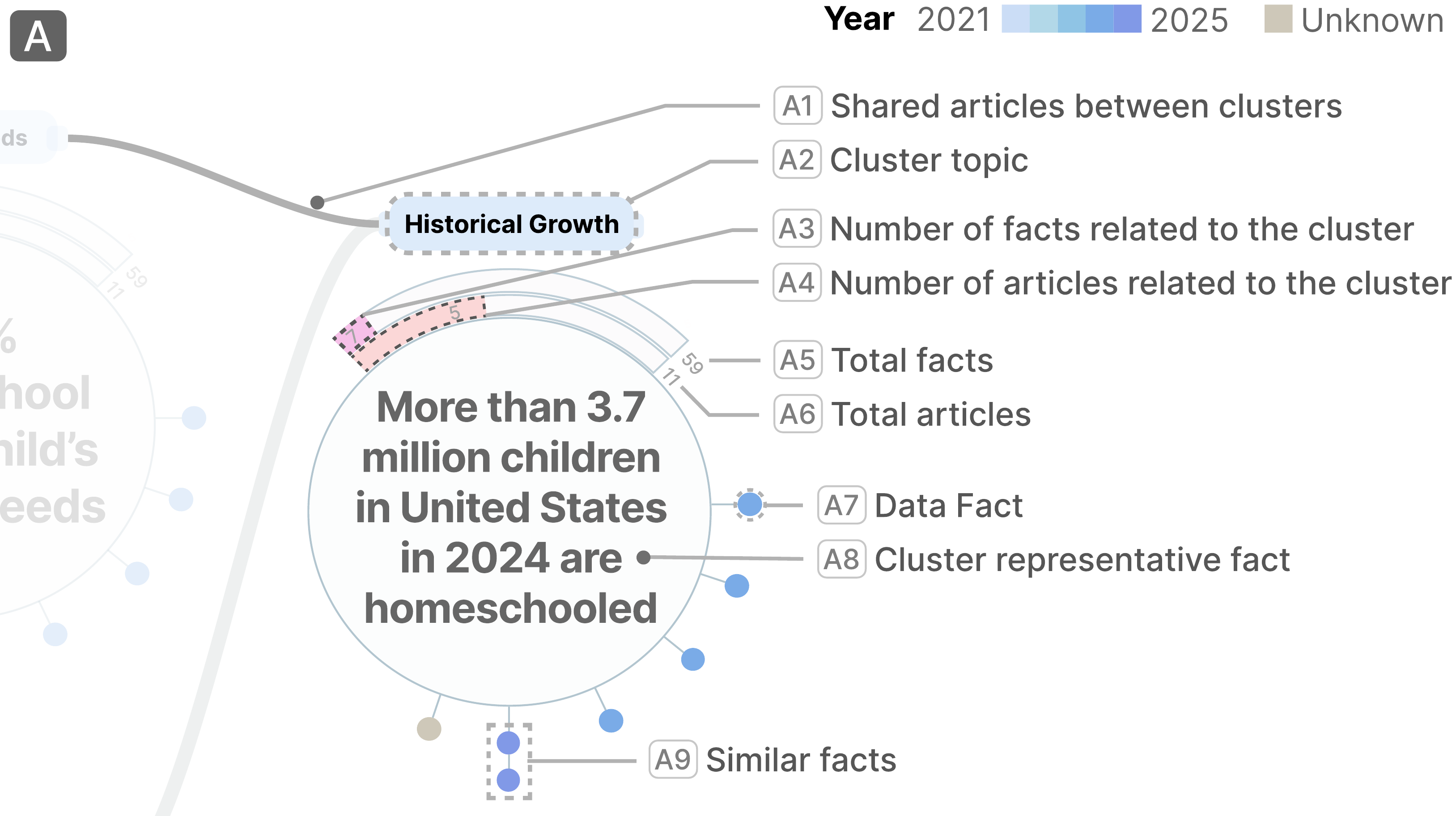}
\caption{%
Thematic Circle represents one cluster of facts, with the cluster representative fact (A8) displayed in the inner circle, and individual data facts (A7) positioned in the bottom half of the outer area, where each circle is color-coded with publication year of its associated article.
}
\vspace{-1em}
\label{figs:clusterencode}
\end{figure}

\subsubsection{Thematic Circle}
The \textbf{Thematic Circle} (Fig.~\ref{figs:clusterencode}) is designed to visualize details of a cluster, including a concise topic and related fact details, allowing users to gain a quick summary of a cluster.
% As shown in Fig.~\ref{figs:clusterencode}\inlineimg{A2}, 
The concise \textit{cluster topic}, displayed at the top (Fig.~\ref{figs:clusterencode}\inlineimg{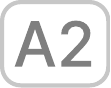}), indicates the main theme of the cluster. 
The circle (Fig.~\ref{figs:clusterencode}\inlineimg{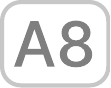}), fixed in size, encloses the text of the cluster’s most representative fact.
The dots (Fig.~\ref{figs:clusterencode}\inlineimg{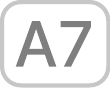}) positioned in the bottom half of the circle display individual data facts in the cluster.
Each dot is color-coded by the publication year of its source article (\req{R5}), allowing users to quickly assess the timeliness of the information.
As shown in Fig.~\ref{figs:clusterencode}\inlineimg{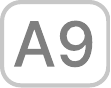}, 
connected dots indicate 
similar facts identified through the merging process described in Section~\ref{sec:factmerging}.
The two radial bars indicate the number of facts (Fig.~\ref{figs:clusterencode}\inlineimg{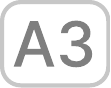})
and the number of contributing articles
(Fig.~\ref{figs:clusterencode}\inlineimg{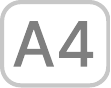}) related to the cluster, enabling easier comparison between clusters (\req{R1}).

The design balances overview and detail to satisfy \req{R1}. 
While displaying the details of all the facts could be overwhelming~\cite{eppler2004concept}, 
showing only a brief topic overview, such as a word cloud of keywords, is often ineffective for interpretation~\cite{harris2011word}.
Our design adopts a focus+context strategy~\cite{cockburn2009review} that balances overview and detail.
Also, inspired by the \textit{Circle Map} used in structured brainstorming~\cite{hyerle2008thinking}: the inner circle conveys the main theme, while the outer ring presents supporting details.

A link (Fig.~\ref{figs:clusterencode}\inlineimg{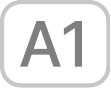}) between two clusters indicates the number of shared articles between them (\req{R2}), where those articles contain information relevant to multiple topics.
The thickness of each link reflects the number of shared articles.

\subsubsection{Filter Widget}\label{sec:fliterwidgets}
As shown in Fig.\ref{fig:teaser}\inlineimg{A.pdf}, \textbf{Filter Widget} helps to manage visual complexity and support diverse exploration needs. 
The filters include: 
(1) \textbf{Relevance-based topic filtering}, a single-thumb slider, which limits the number of displayed \textit{Thematic Circles} (six by default) based on their relevance to the query (Section~\ref{sec:topicclustering}); 
(2) \textbf{Shared article connection filtering}, a range slider, which adjusts the visibility of inter-topic links based on the number of shared articles to reduce clutter and support targeted exploration; and 
(3) \textbf{Fact visibility toggling}, which lets users show or hide fact dots to maintain a cleaner view.
Based on the applied filters, the summary panel (Fig.~\ref{fig:teaser}\inlineimg{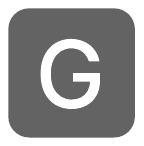}) updates to summarize both the statistics of the current story and the overall distribution of articles and facts, providing a concise snapshot of the information landscape.

\subsubsection{Interactions}
\toolName{} enables rich interactions to allow users to smoothly explore the overview of the story.
Users can hover over a data fact dot (Fig.~\ref{figs:clusterencode}\inlineimg{A7.pdf}) to reveal the detailed fact panel (Fig.~\ref{fig:teaser}\inlineimg{B.pdf}) displaying the detailed fact, its source, and publication year (\req{R5}).
Furthermore, each favicon  (Fig.~\ref{fig:storyview}\inlineimg{Ag3.pdf}), which indicates the source article, is clickable and redirects users to the original article (\req{R5}).
Hovering over the related articles bar (Fig.~\ref{figs:clusterencode}\inlineimg{A4.pdf})  reveals the list of the articles contributing to the cluster as shown in Fig.~\ref{fig:teaser}\inlineimg{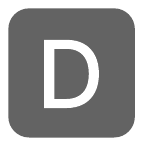}.
Similarly, hovering over the radial bar for related facts (Fig.~\ref{figs:clusterencode}\inlineimg{A3.pdf}) displays a panel showing the count of related facts (Fig.~\ref{fig:teaser}\inlineimg{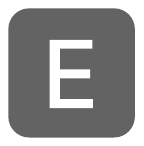}).
Hovering over a link (Fig.~\ref{figs:clusterencode}\inlineimg{A1.pdf}) reveals the shared articles panel (Fig.~\ref{fig:teaser}\inlineimg{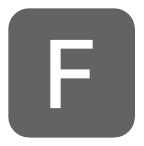}), which lists the information about the shared articles (\req{R2}), including their titles, publication years, and sources.
Users can also toggle the articles panel (Fig.~\ref{fig:teaser}\inlineimg{C.pdf}) by clicking on the bottom-right arrow button as shown in Fig.~\ref{fig:teaser}.
It shows all the articles contributing to the story, including their titles, snippets, sources, and published years (\req{R5}).

\vspace{-1em}
\subsection{Story View}
The \textbf{Story View} (Fig.~\ref{fig:storyview}\inlineimg{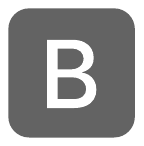}) 
allows users to move from the overview to detailed exploration of the data facts of a topic, helping them understand the quantitative evidence. 
This consists of a \textbf{Narrative Unit} (Fig.~\ref{fig:storyview}\inlineimg{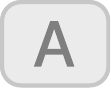}), which provides a detailed story with a chart, and a zoomed \textit{Thematic Circle} (Fig.~\ref{fig:storyview}\inlineimg{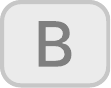}), which presents the top three most relevant data facts as text within circles (Fig.~\ref{fig:storyview}\inlineimg{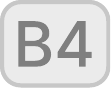}), rather than showing only the most representative fact.

\subsubsection{Narrative Unit}\label{sec:narrativeunit}
\begin{figure*}[t] % Use figure* for a two-column figure, [t] for top of the page
  \centering % More compact than using \begin{center} ... \end{center}
   \setlength{\abovecaptionskip}{0.1em} 
  \includegraphics[width=\textwidth]{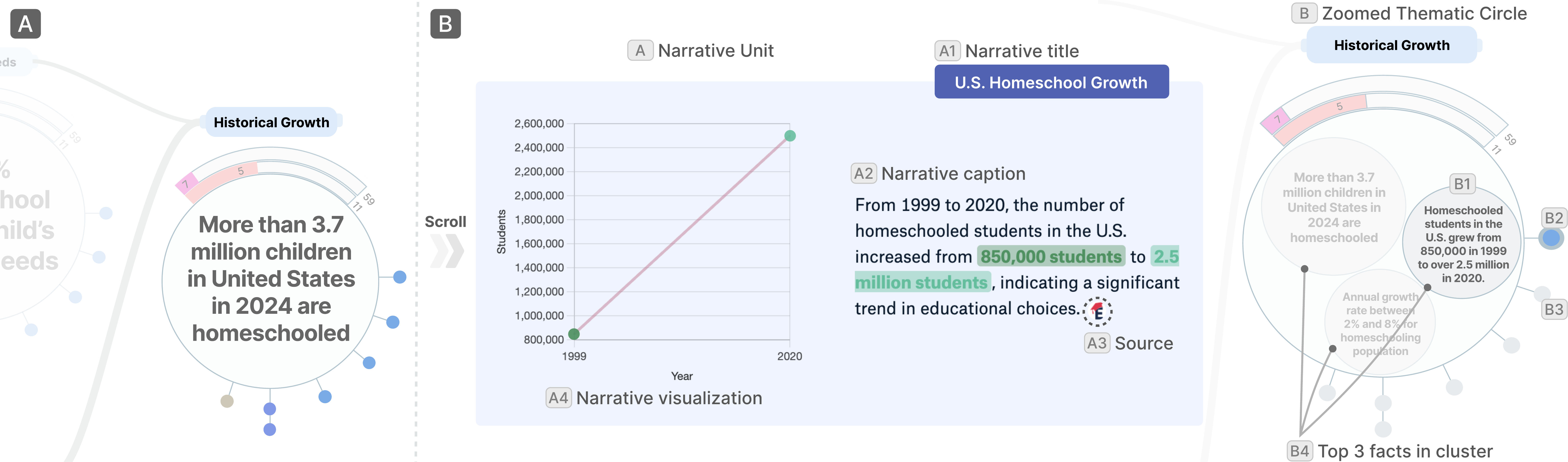} % Use \textwidth for spanning across two columns
   \caption{%
  Scrolly flow of \toolName{}. The system transitions from \textbf{Thematic Overview} (A) to \textbf{Story View} (B) when scrolling down. (B) illustrates the narrative related to U.S. homeschool growth, where the corresponding fact is highlighted in the outer area of the zoomed thematic circle. Since this narrative is derived from one of the top 3 facts in the cluster, the related fact is also highlighted within the zoomed thematic circle.
  }
  \vspace{-1em} 
  \label{fig:storyview}
\end{figure*}

The \textbf{Narrative Unit}, shown in Fig.\ref{fig:storyview}\inlineimg{figs/Ag.pdf}, summarizes a \textit{merged fact set} derived in Section~\ref{sec:factmerging}, combining a text description (Fig.\ref{fig:storyview}\inlineimg{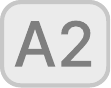}) with a chart (Fig.\ref{fig:storyview}\inlineimg{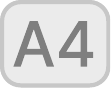}) in detail (\req{R4}).
The \textbf{narrative title}, shown in the top right of the \textit{Narrative Unit} (Fig.~\ref{fig:storyview}\inlineimg{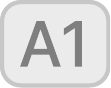}), conveys the core message of the \textit{merged fact set}, generated by prompting an LLM to give a concise title of the \textit{merged fact set}.

The \textbf{narrative caption}, a text with highlights, shown in  Fig.~\ref{fig:storyview}\inlineimg{figs/Ag2.pdf}, expresses the facts, in a \textit{merged fact set}, in natural language to form a meaningful story.
The caption highlights key numerical values, applying consistent colors to match the data values across both the caption and the visualization (\req{R4}).
The favicons at the end of the text (Fig.~\ref{fig:storyview}\inlineimg{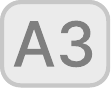}) indicate the source articles for the narrative (\req{R5}).
This \textit{narrative caption} is generated during the \textbf{Narrative Generation} phase of our framework (Fig.~\ref{fig:framework}\inlineimg{figs/B.pdf}), as HTML code, using an LLM to highlight and summarize all the facts within a \textit{merged fact set} and produce a concise, human-readable story.

The \textbf{narrative visualization}, shown in Fig.~\ref{fig:storyview}\inlineimg{figs/Ag4.pdf}, illustrates the underlying data of a \textit{merged fact set} (\req{R4}). 
It can take the form of bar, pie, line, isotype, range/dumbbell, or text, which are the commonly-used charts in news and data-driven articles, as also noted in prior research\cite{shi2020calliope, wang2019datashot}.
The \textbf{Visualization Preparation} phase of our framework (Fig.~\ref{fig:framework}\inlineimg{figs/B.pdf}) recommends the chart type using an LLM, informed by prior work~\cite{wang2023llm4vis} that uses LLMs to recommend appropriate visualizations.

\subsubsection{Interactions}
\rvm{\toolName{} provides a scroll-driven narrative interface that implements scrollytelling techniques~\cite{oesch2022scrolling} to support both linear and non-linear exploration (\req{R3}).}

To enable a smooth story flow among different \textit{Narrative Units}, we arrange them in a coherent sequence at two levels (\textbf{Narrative Organization} in Fig.~\ref{fig:framework}\inlineimg{figs/B.pdf}): across clusters and within each cluster, as each cluster contains one or more \textit{Narrative Units}.
First, the inter-cluster ordering follows the inverted pyramid paradigm for engaging news reporting~\cite{Potker01012003} by placing the most relevant clusters first based on the clusters' relevance scores (Section~\ref{sec:topicclustering}).
Second, the intra-cluster ordering uses an LLM to arrange \textit{Narrative Units} within a cluster, producing a coherent sequence with smooth transitions between \textit{Narrative Units}.
This layered approach transforms all \textit{Narrative Units} into a coherent data story.

\textbf{Linear exploration.}
As users scroll down the page, \toolName{} progressively unfolds the story following the above-mentioned order.
When users begin scrolling on the \textit{Thematic Overview}, \toolName{} smoothly transitions to the first \textit{Story View} (Fig.~\ref{fig:storyview}\inlineimg{figs/B.pdf}) using an animated pan and zoom to focus on the first \textit{Thematic Circle}, while keeping other circles subtly visible in the background, following the \q{Pan-And-Zoom} scrollytelling technique~\cite{oesch2022scrolling} (\req{R3}).
The zoomed \textit{Thematic Circle} highlights facts associated with the current \textit{Narrative Unit} (Fig.~\ref{fig:storyview}\inlineimg{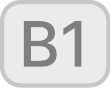}, \inlineimg{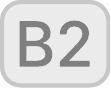}), while dimming others (Fig.~\ref{fig:storyview}\inlineimg{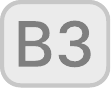}) (\req{R5}).
\rvm{As users scroll to the next \textit{Narrative Unit} within the same cluster, the zoomed \textit{Thematic Circle} animates to highlight the facts related to the new \textit{Narrative Unit}, smoothly transitioning between visual states (\req{R3}). 
For example, when scrolling, we animate the \textit{Thematic Circle} from Fig.~\ref{figs:casestudy}\inlineimg{figs/B.pdf} to Fig.~\ref{figs:casestudy}\inlineimg{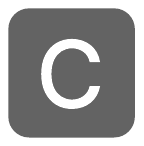}, updating the highlights such as most representative facts (Fig.~\ref{figs:casestudy}\inlineimg{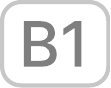}) and fact dots (Fig.~\ref{figs:casestudy}\inlineimg{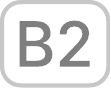}).
When users scroll past the last narrative of the current cluster to a new cluster, the system pans and zooms to its \textit{Thematic Circle} and unfolds the \textit{Narrative Units}.}

\textbf{Non-linear exploration.}
In addition to scroll-driven progression, users can jump to any cluster by clicking its \textit{Thematic Circle} (Fig.~\ref{figs:clusterencode}).
The system navigates to that cluster’s first \textit{Narrative Unit}, from which users continue scrolling to reveal the remaining \textit{Narrative Units} as in linear exploration.
Users can return to the \textit{Thematic Overview} by clicking the outside (Fig.~\ref{figs:casestudy}\inlineimg{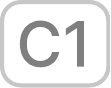}) of the \textit{Thematic Circle}.

\section{Use Case}

In this section, we present a use case scenario where Bob, 
a university student and avid news reader, is analyzing TikTok’s global trends for a media studies assignment. 
Rather than looking for subjective takes, 
Bob aims to uncover objective, quantitative facts 
about the TikTok’s worldwide impact.

\begin{figure*}[!htbp] % Use figure* for a two-column figure, [t] for top of the page
  \centering % More compact than using \begin{center} ... \end{center}
   \setlength{\abovecaptionskip}{0.1em} 
  \includegraphics[width=\textwidth]{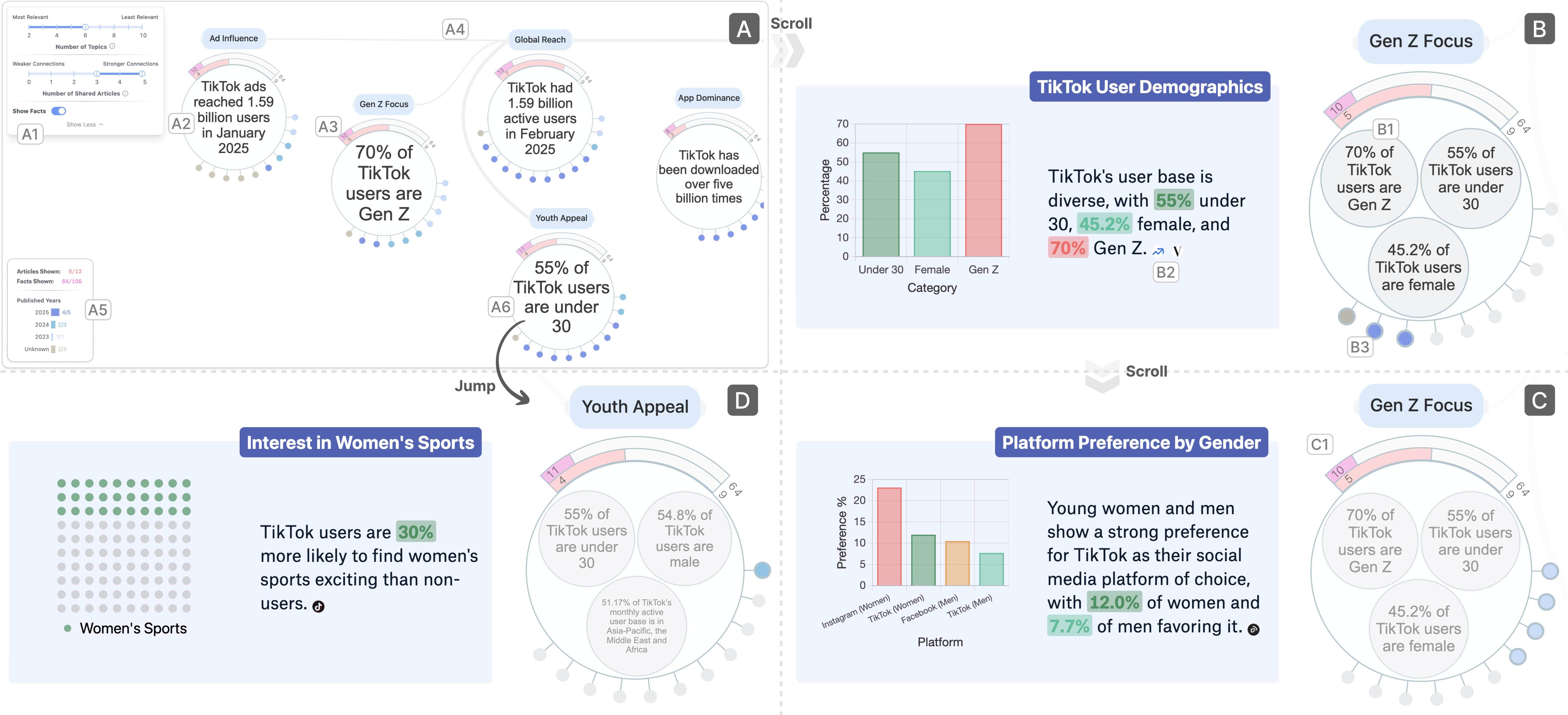} % Use \textwidth for spanning across two columns
   \caption{%
  \toolName{} generated story for a query \q{TikTok trends worldwide} showcasing: (A) the Thematic Overview; (B--C) the continuous scroll-down flow from (A) to (C) to explore the story; and (D) jumping to the story in the \textbf{Youth Appeal} theme by clicking the thematic circle (A6) in (A). 
  }
  \vspace{-1em} 
  \label{figs:casestudy}
\end{figure*}

Bob discovers our system, \toolName{}, and enters the query \q{TikTok trends worldwide}.
Instead of returning a flat list of articles, the system expands the query with variations, retrieves relevant articles, and extracts quantitative facts.
Within moments, \toolName{} transforms his query into a rich, interactive visual overview (Fig.~\ref{figs:casestudy}\inlineimg{A.pdf}).
From 12 retrieved articles, the system identifies 106 distinct facts and organizes them into thematic clusters such as Global Reach, Gen Z Focus, App Dominance and others. 
Initially, it presents 64 facts drawn from the six most relevant clusters, sourced from 9 of the 12 articles as displayed in summary panel (Fig.\ref{figs:casestudy}\inlineimg{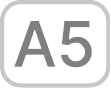}), giving Bob a first glance of the most relevant themes.
Yet, he wants to dig deeper into the evidence.

Eager to explore the story behind the clusters,
he scrolls down, and the system smoothly pans and zooms into the \textit{Gen Z Focus} cluster from the overview.
The story begins with \textit{TikTok User Demographics} (Fig.\ref{figs:casestudy}\inlineimg{B.pdf}), where Bob learns that 70\% of TikTok’s users are from Generation Z.
The highlighted circles (Fig.\ref{figs:casestudy}\inlineimg{B1.pdf}) show that this story draws on three key facts, while the favicons (Fig.~\ref{figs:casestudy}\inlineimg{B2.pdf}) and highlighted dots (Fig.~\ref{figs:casestudy}\inlineimg{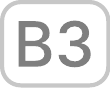}) show that the story draws on two articles and three data facts.
Wanting to check the credibility of the data, Bob hovers over the fact dot (Fig.~\ref{figs:casestudy}\inlineimg{B3.pdf}) to see the concise fact details as in the Fig.~\ref{fig:teaser}\inlineimg{B.pdf} panel,
and clicks on the favicon to trace it back to the original sources, confirming their credibility.

As he scrolls further, new narratives emerge, for example, on \textit{Platform Preference by Gender} (Fig.~\ref{figs:casestudy}\inlineimg{C.pdf}). 
In the background, the \textit{Thematic Circle} dynamically updates, highlighting the facts relevant to each unfolding story. 
This scrollytelling flow helps Bob stay connected to the overall story while drawing him deeper into the evidence like user growth, download trends, revenue growth and many more fulfilling his goal of finding quantitative facts.
After following the initial clusters, Bob grows curious about what else he might uncover.
With a simple click outside the thematic circle (Fig.\ref{figs:casestudy}\inlineimg{C1.pdf}), he moves back into the overview (Fig.\ref{figs:casestudy}\inlineimg{A.pdf}).
Using the filter controls (Fig.\ref{figs:casestudy}\inlineimg{A1.pdf}), 
he expands the number of visible topics and pays his attention to the themes with more facts and broader article coverage, 
easily identified by the radial bars (Fig.~\ref{figs:casestudy}\inlineimg{A3.pdf}).

His curiosity leads him to the \textit{Youth Appeal} theme. 
Clicking on the \textit{Thematic Circle} (Fig.~\ref{figs:casestudy}\inlineimg{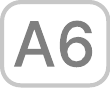}), he jumps into a related narrative, where he discovers that 30\% of TikTok users express interest in women’s sports (Fig.~\ref{figs:casestudy}\inlineimg{D.pdf}).
With each new discovery, Bob becomes increasingly impressed by \toolName{}’s ability to distill scattered news coverage into clear, engaging, and objective visual narratives.

\section{Evaluation}

We extensively evaluated \toolName{} through both quantitative evaluation and two user studies.
\rvm{The quantitative evaluation measured the accuracy of results at different stages of our framework. 
The two user studies provided a comprehensive evaluation of \toolName{}, with the first study assessing usability and effectiveness, and the second study evaluating ecological validity and real-world utility.}
More details related to the studies are provided in the supplementary materials.

\begin{figure*}[!t] % Use figure* for a two-column figure, [t] for top of the page
  \centering % More compact than using \begin{center} ... \end{center}
  \setlength{\abovecaptionskip}{0.1em} 
  \includegraphics[width=\textwidth]{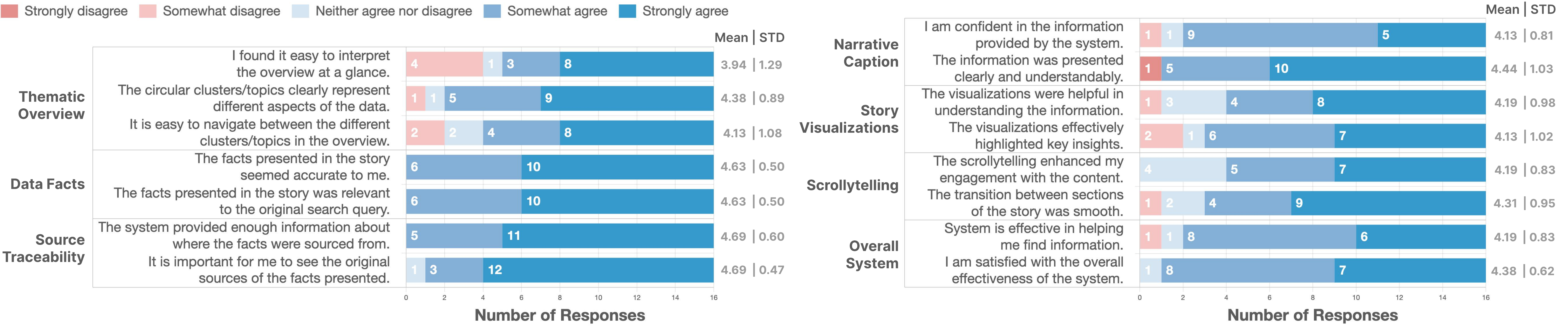} % Use \textwidth for spanning across two columns
   \caption{%
The effectiveness of \toolName{}. Responses cover aspects of the Thematic Overview, including interpretability, theme clarity, and navigation; Data Facts with their perceived accuracy and relevance; Source Traceability for factual grounding; Story View in terms of Narrative Caption clarity and confidence, usefulness of Story Visualizations, Scrollytelling engagement and transitions, and Overall System effectiveness and satisfaction. 
  }
  \vspace{-1.5em} 
\label{figs:effectiveness}
\end{figure*}

\vspace{-1em} 
\subsection{Quantitative Evaluation}
\paragraph*{\textbf{Procedure}}\label{humanevalprocesure}
We generated 
four data stories for topics in distinct domains (\eg \q{Tiktok trends worldwide}) using \toolName{}.
In total, 399 extracted data facts
from 36 articles, 
producing 196 \textit{merged fact sets} along with their associated \textit{narrative captions}
and \textit{visualization recommendations}.
Two co-authors independently reviewed and coded each story without prior discussions to maintain objectivity.
They examined five key aspects:
the accuracy of \textit{fact content} and \textit{data points} from the \textit{Data Fact Extraction} stage;
the quality of \textit{merged fact sets} from the \textit{Fact Organization} stage; 
and the quality of \textit{narrative captions} and \textit{visualization recommendations} from the \textit{Visual Storytelling} module.
To evaluate the accuracy of \textit{fact content} and \textit{data points},
the coders manually compared each extracted fact with its source article.
A \textit{fact content} was marked as correct if its meaning and context were faithfully retained, and a \textit{data point} was marked correct if it exactly matched the value in the source.
The quality of \textit{narrative captions} was marked as correct if they faithfully reflected the underlying data facts. A \textit{merged fact set} was marked as correct if the underlying data facts logically fit together. A \textit{visualization recommendation} was marked as correct if it effectively represented the extracted data; otherwise, coders provided their own recommendation.
\paragraph*{\textbf{Results}}\label{humanevalresults}
We adopted accuracy (i.e., the proportion of correct items in all extracted items) as the primary evaluation metric. 
\toolName{} achieved 97.2\% accuracy in both \textit{fact content} and \textit{data points},
95.9\% accuracy in \textit{merged fact sets}, 
92.3\% accuracy in \textit{narrative caption},
and 77\% accuracy in \textit{visualization recommendations}.
These results demonstrate \toolName{}'s effectiveness in accurately extracting structured data from unstructured text and the ability to
connect fragmented facts into coherent stories leveraging the LLMs.

The failure cases in \textit{Data Fact Extraction} stage arose from the difficulties in interpreting comparative expressions (\eg \q{A is 30\% more than B}) to derive the value of one item from the other.
The \textit{Fact Organization} failures arose primarily in cases requiring complex reasoning, 
such as when some facts shared the same context (\eg year, region) while some of them included an additional qualifier (\eg \q{among men}), leading to incorrect merges.
The visualization recommendation failures were mostly with \textit{Narrative Units} that contained a single data point. 
For example, a text visualization was recommended for \q{About 57\% of TikTok users discover brands through social media}, possibly because of the approximate value (\q{about 57\%}) was treated as plain text, whereas coders expected an isotype chart to better highlight the proportion.
We anticipate the presence of such failure cases to decrease as LLMs become better at reasoning about data context.

\vspace{-1em}
\subsection{User Studies}

\rvm{We conducted two user studies with the same participants under controlled and non-controlled conditions.
The \textbf{Initial Perception Study} adopted a controlled design
with a pre-generated data story using \toolName{} for all participants,  ensuring consistent evaluation of usability and effectiveness, and enabling a fair comparison with the AI-powered search engine Perplexity.
Informed by the findings from the initial study, we updated \toolName{} by adding interactive filters (Fig.~\ref{fig:teaser}\inlineimg{A.pdf}) prior to conducting the second study.
The \textbf{Open-Ended Query Exploration Study} adopted a non-controlled design, allowing participants to pursue self-directed queries to assess ecological validity and real-world utility.}
All the studies were approved by our university’s Institutional Review Board.

\subsubsection{\textbf{Initial Perception Study}} 

\begin{figure}[!t]
\centering
 \setlength{\abovecaptionskip}{0.1em} 
\includegraphics[width=\columnwidth]{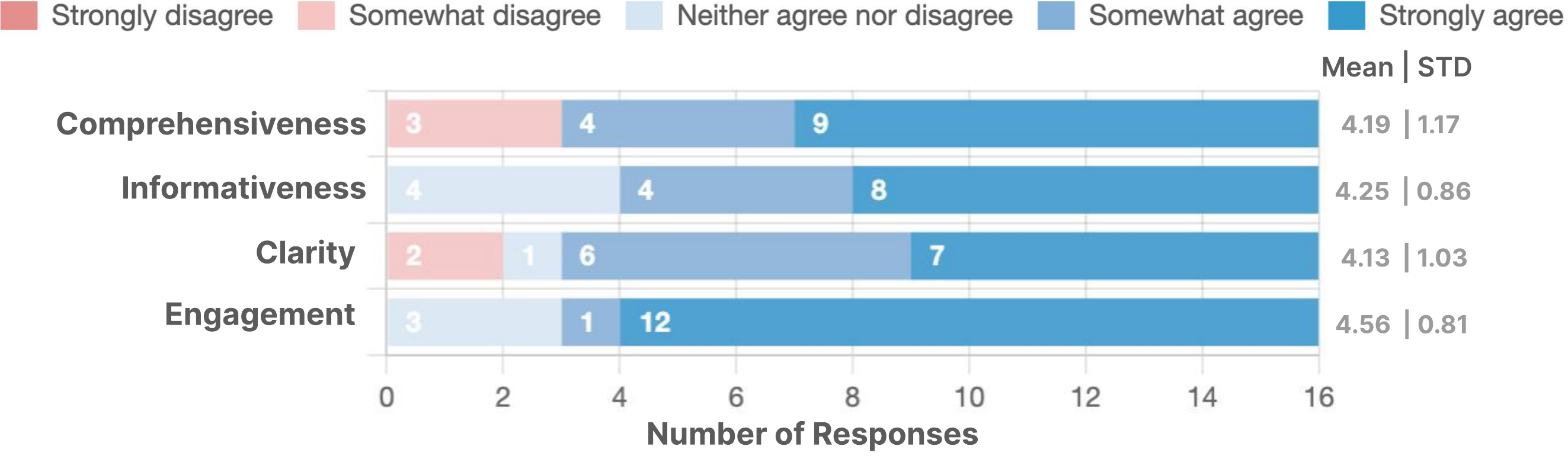}
\caption{% 
The comparison between \toolName{} and Perplexity.}
\vspace{-1.5em} 
\label{figs:comparison}
\end{figure}

This study evaluates the usability and effectiveness of the initial system and compares it with
the baseline AI-powered search engine Perplexity.
\paragraph*{\textbf{Participants}}
We recruited 16 participants (P1–P16) from two universities (8 females, 8 males,
\ensuremath{\mu_{\text{age}}}= 25.75 years, \ensuremath{\sigma_{\text{age}}}= 3.75). 
All participants reported daily use of web search and AI-based search tools, primarily ChatGPT and Perplexity.
Additionally, 13 participants were familiar with scrollytelling.
\paragraph*{\textbf{Procedure}}
The studies were conducted through video calls
where participants accessed an online version of \toolName{},
while sharing their screens. 
We first collected participants’ demographic information and then introduced them to the system through a data story (\q{Does AI lead to new jobs or job displacements?}), which they explored to become familiar with its features.
Then, they were tasked to examine a data story (\q{Is homeschooling preferred by people?}) using a think‑aloud protocol,
freely navigating while ensuring full coverage. 
Next, they used Perplexity to search the same query and reviewed its results. 
Finally, participants completed a post‑study questionnaire (5‑point Likert) on \toolName{}’s effectiveness (Fig.~\ref{figs:effectiveness}), rated usability with the System Usability Score (SUS)~\cite{brooke1996sus}, compared \toolName{} with Perplexity (Fig.~\ref{figs:comparison}), and joined semi‑structured interviews. 
The whole study lasted about 50 minutes.

\paragraph*{\textbf{Results}}
Fig.~\ref{figs:effectiveness} summarizes the questionnaire responses.
As it shows, participants rated \toolName{} favorably in terms of its visualization design (Thematic Overview and Story Visualization), the quality of the data facts (Data Facts and Source Traceability), and the clarity and coherence of the stories (Narrative Caption and Scrollytelling).
They also found the overall system effective and helpful. 
\toolName{} received a SUS score of 73.3, which falls within the good range~\cite{bangor2009determining}.

\underline{\textit{Thematic Overview.}}
Participants found the thematic overview was easy to interpret
\ratingstats{3.94}{1.29}, effective for representing different aspects of the information \ratingstats{4.38}{0.89}, 
and smooth for navigating between clusters \ratingstats{4.13}{1.08}. 
These ratings support our overview first approach.
P3 valued \q{seeing the big picture} beyond a simple Google search, and P10 appreciated the links and timeline cues.
We further observed that most participants used hovering features to see more details of the facts and the articles. 
However, while participants appreciated the information richness in the \textit{Thematic Overview}, 
others, especially P2 and P11, found the overview somewhat complex and suggested adding more controls, such as toggling connections on or off (P2).
This led us to integrate filter widget into the updated system.

\underline{\textit{Facts and Source Traceability.}}
Participants valued the concise, accurate facts \ratingstats{4.63}{0.50} and strong source traceability \ratingstats{4.69}{0.60}, with links enabling verification \ratingstats{4.69}{0.47}.
P11 appreciated that \q{the facts are present in short and crisp format.}
We observed that the participants used source links not only for verification but also to explore broader context, enhancing the trust and engagement.

\underline{\textit{Story View.}}
Participants found the story-like format effective for organizing information and navigating between narratives, with smooth animated transitions \ratingstats{4.31}{0.95} and engaging scrollytelling \ratingstats{4.19}{0.83}. 
Most did not feel lost, though some suggested adding a minimap-style navigator, similar to those in IDEs, to show their current position in the story. 
Narrative captions were rated clear \ratingstats{4.44}{1.03} and coherent \ratingstats{4.13}{0.81}, 
although a few (\eg P15) wanted more background details.
Visualizations were valued for aiding comprehension \ratingstats{4.19}{0.98} and highlighting key insights \ratingstats{4.13}{1.02}.
Meanwhile, some participants noted failure cases, including duplicated narratives caused by hallucination, 
incoherent narratives (\eg where different facts described the same issue but reported conflicting numerical values were merged into a single narrative), 
and instances where the data visualizations did not fit the content.
During the interviews, P3 suggested incorporating images and using charts from the original articles, while P5 recommended retrieving more data to make the visualizations richer.

\underline{\textit{Overall System.}}
Overall, \toolName{} was seen as effective for gaining a comprehensive view \ratingstats{4.19}{0.83} with high satisfaction \ratingstats{4.38}{0.62}. 
Most participants valued its ability to condense scattered quantitative information into clear narratives and indicated they would use it for research (P8), academic work (P12), and sharing stories (P13).

\textbf{Comparison.}
Participants compared \toolName{} with Perplexity in terms of comprehensiveness, informativeness, clarity, and engagement as shown in Fig.~\ref{figs:comparison}. 
The interviews revealed two clear strengths. 
First, participants valued the integrated overview of the topics in \toolName{}, 
which let them see the full topical landscape before drilling into specifics.
As P3 explained, \q{Compendia shows me the whole picture, so I know where to start.}
Second, they valued the system’s storytelling approach, where we combine data visualizations with the text, as the visuals helped them see trends and make comparisons that pure text could not. 
P10 remarked, \q{Perplexity is more like an essay answer but this combines visuals with text.}
However, not all the participants preferred this style. 
For example, P15, who mentioned a preference for reading text, leaned towards text-only summaries.
Furthermore, P2 suggested adding a brief summary paragraph, similar to the opening paragraph in Perplexity’s results.

\subsubsection{\textbf{Open-Ended Query Exploration Study}}
\label{sec:open-ended}
This study examines the real-world use of \toolName{} by letting the participants try their own open‑ended queries.

\paragraph*{\textbf{Participants}} \rvm{We recruited the same 16 participants (P1--P16) involved in the initial perception study to join this study.}

\paragraph*{\textbf{Procedure}}
The studies were conducted through video calls
where participants accessed an online version of \toolName{},
while sharing their screens.
Participants were asked to formulate a search query of their interest and searched in \toolName{}. 
Once the story was generated, they explored it without any guidance or restrictions.
\rvm{We observed participants' interactions throughout the study.}
Finally, participants completed a post-study questionnaire and semi-structured interviews. 
The sessions lasted about 30 minutes.
They received \$25 for their time participating in both user studies.

\paragraph*{\textbf{Results}}
\rvm{We summarize the study results in terms of the generalizability of \toolName{}, information foraging and sensemaking strategies, and improvement suggestions.}

\rvm{\underline{\textit{Generalizability across domains and task types.}}
All participants successfully performed a self-directed query using \toolName{}.
The queries spanned multiple domains, including 
real estate (\eg \q{What are the property prices for Sydney 2025?} – P10), 
demographics (\eg \q{Marriage statistics in Singapore} – P4),
healthcare (\eg \q{The distribution of AIDS patients in Sweden} – P12),
finance (\eg \q{What are the market shares of global pharmaceutical companies?} – P3),
and architecture (\eg \q{Typical Gothic architectural style in Notre Dame de Paris?} – P11).
These queries also reflected multiple information-seeking task types, such as 
comparative analysis (\eg \q{BYD car sales compared to Tesla car sales in the recent years} – P1), 
trend or temporal analysis (\eg \q{AI usage trends} – P9), 
forecasting (\eg \q{Gold prices predictions by different firms} – P6), 
and descriptive statistical overview (\eg \q{World population by country} – P15).
Together, these findings suggest broader applicability of \toolName{} across different domains and task types in real-world scenarios.}

\rvm{\underline{\textit{Information foraging and sensemaking strategies.}}
All the participants agreed that the resulting information generally matched their expectations. 
For example, P7 mentioned \q{One of my friends shared this [resulted story] Harvard funding freeze with me a few weeks ago.}
We observed three recurring information foraging strategies that supported effective sensemaking: }

\rvm{\textit{\textbf{Overview-first exploration.}} 
Most of the participants began by looking at the \textit{Thematic Overview} to assess topic coverage before proceeding to detailed stories; as P9 explained, \q{The circular layout [thematic overview] helped me pick up on some high-level AI trends I didn’t thought first.}
Then, they followed the story in the given sequence by scrolling.}

\rvm{\textit{\textbf{Back-and-forth navigation and iterative sensemaking.}} 
Rather than examining the entire overview at once, some participants (P4 and P14) moved iteratively between the \textit{Thematic Overview} and the detailed \textit{Narrative Units} to refine their understanding as new information emerged.
For instance, P4 initially navigated to detailed narrative related to the Divorce Trends, which appeared top left of the overview, and then returned to explore other connected topics.}

\rvm{\textit{\textbf{Filtering for sensemaking.}}
All participants agreed that the filters were easy to understand and useful for information foraging. 
Beyond surface-level navigation, several participants used filtering as an active sensemaking strategy to identify salient patterns and relationships. 
For example, P3, who explored market shares of global pharmaceutical companies, first filtered by the most relevant topics, which helped them quickly surface dominant players, noting that \q{Pfizer is dominating this.} 
Similarly, P4 used the shared-articles filter to examine cross-article connections, which prompted reflection on broader patterns, observing that \q{These links suggest educated people tend to have fewer children.}}

\rvm{\underline{\textit{Improvement suggestions.}}
During the study, some participants encountered minor issues that highlight opportunities for further refinement of the system in future work.}

\rvm{\textit{\textbf{Handling outdated and time-sensitive information.}}
Some participants observed outdated results for future-oriented predictions.
For example, P6 searched for gold price predictions and expected forecasts for the coming years.
While the results included such forecasts, 
they also surfaced predictions from older articles that were no longer valid at the time of exploration (\eg a 2023 article forecasting prices for 2024, which was outdated in 2025).
This limitation primarily occurs in cases involving time-sensitive predictive information, which the current system does not explicitly distinguish.
As future work, this issue could be addressed by incorporating explicit temporal reasoning into the framework (in \textit{Data Validation}).}

\rvm{\textit{\textbf{Redundancy and narrative coherence.}}
Some participants noticed somewhat similar content appearing across different clusters (P12), as well as occasional disjointed story flow (P8).
These issues primarily stemmed from phrasing variations across sources that express equivalent facts differently, causing them to be clustered into separate groups, as well as the high diversity of extracted facts, which can disrupt smooth narrative transitions.
Although relatively infrequent, these cases highlight opportunities for future work to improve clustering and narrative organization and
are likely to diminish as LLM context windows continue to expand.}

\section{Discussion}

This section discusses key lessons learned from generating data stories using unstructured text and outlines limitations and future work.

\vspace{-2em}
\rvm{\subsection{Lessons Learned}}

\rvm{\textbf{LLMs for handling unstructured text:}
Systematically transforming unstructured information into structured narratives requires effective information retrieval and extraction.
Transformer-based retrieval models (\eg BERT) often require domain-specific fine-tuning, limiting cross-domain applicability, 
while rule-based and statistical extraction methods (\eg NER) struggle with linguistic ambiguity and stylistic or structural variation.
In contrast, LLMs offer a more flexible alternative, as they can leverage contextual reasoning even in the absence of explicit task-specific training.
However, LLMs are inherently unreliable, and their effectiveness in data storytelling relies on the integration with a broader, more robust processing framework.
Several strategies in our processing framework were helpful for reliability.
Text chunking and parallel processing enable handling articles that exceed an LLM’s context window, by splitting text into manageable segments that are processed independently and later aggregated.
Structured output and schema enforcement constrain model outputs to predefined formats, improving consistency and downstream usability.
Validation and retry logic ensure that extracted results satisfy schema and quality requirements.
Finally, prompt engineering plays a central role, with clear instructions and few-shot examples guiding model behavior without requiring model fine-tuning.}

\rvm{\textbf{Challenges in LLM-based systems:}
While LLMs offer flexibility in handling unstructured text, they also introduce practical challenges when it comes to larger context, specific tasks, and time efficiency. 
Scalability remains a primary concern, particularly in the later stages of the framework (\eg \textit{Fact Organization}), where all extracted facts must be considered at once and chunking is not feasible.
For example, our experiments with LLM-based clustering yielded poor and unstable results (\eg missing or hallucinated facts), due to both the larger context and the inherently unsupervised nature of the task.
This prompted us to adopt a more robust machine learning approach for clustering. 
In practice, the system can effectively manage around 300 individual facts per story, 
constrained by output context length limits (especially in merging and validation) that may result in incomplete output structures.
We believe these constraints will eventually be resolved with the rapid context length expansion of modern LLMs~\cite{ding2024longrope}.
Another drawback of using LLMs is the latency. 
This limitation is not unique to our system; 
most advanced research agents that autonomously search, analyze, and synthesize information from the Internet, such as DeepResearch\footnote{\url{https://openai.com/index/introducing-deep-research/}}, also require longer processing time than regular LLM-based chat.
As LLM architectures continue to evolve, we expect the future models to achieve lower latency with high accuracy.}

\rvm{\textbf{Support personalized data stories:} 
Throughout the user studies, we observed that participants employed distinct sensemaking strategies and expressed different information needs when interpreting the presented content.
For example, P3 began by filtering for only the most relevant information, whereas P15 preferred richer contextual details to better understand extracted facts.
These variations highlight the inherently personal nature of sensemaking and suggest future research opportunities in adaptively personalizing data stories based on factors such as users’ interaction histories (\eg previously applied filters), information granularity (\eg preference for detailed contextual information), task framing (\eg comparative analysis), and source preferences.}

\rvm{\textbf{Storytelling for search interfaces:} We believe this work opens new directions for search interfaces, beyond presenting ranked lists such as map-based visualizations~\cite{jaakko2017topicrelevancemap, edward2009resultmaps}, where narrative storytelling can enhance how users engage with and interpret search results.
In our study, participants generally preferred \toolName{} over traditional text-based interfaces like Perplexity, citing its structured presentation and visual storytelling. 
This aligns with emerging trends in search technologies, such as Bing’s Copilot Search\footnote{\href{https://blogs.bing.com/search/2021_03/Microsoft-Bing-delivers-more-visually-immersive-experiences-that-save-you-time}{https://blogs.bing.com/search/2021\_03}}, which are beginning to integrate AI-generated summaries, images, and infographics to give visually immersive search experiences.
A promising avenue for future work is to expand beyond factual content toward nuanced perspectives such as opinions, interpretations, and contested views while preserving clarity and trust.}

\rvm{\subsection{Limitations and Future Work}}
\label{sec:limitations}

\rvm{\textbf{Fact-checking and bias awareness:}
\toolName{} provides source links alongside the extracted facts and generated narratives to support transparency and verification.
However, the system does not independently verify extracted facts or assess framing and bias, which are beyond the scope of this work.
As a result, inaccurate, outdated, or selectively reported statistics in the source articles may still propagate into the generated story, potentially undermining the result quality.
Automated fact-checking and bias assessment remain active research areas~\cite{agunlejika2025ai,guo2022survey}.
Future work could integrate retrieval-based verification, knowledge-graph validation, or claim--evidence reasoning, and 
employ visual cues to explicitly communicate uncertainty or confidence levels.}

\rvm{\textbf{Objectivity--context trade-off:}
To preserve objective grounding, we exclude subjective opinions and interpretations, keeping only minimal context for each fact (e.g., its source sentence).
While this reduces the risk of introducing speculative explanations and makes facts easier to grasp, it might omit qualitative context that helps readers interpret \q{why} quantitative changes occur (\eg policy shifts).
As a result, facts might appear isolated and connections across narratives  could feel less coherent in some cases, where broader contexts help weave a more coherent overall narrative.
Future work could add richer context and carefully scoped interpretive statements to better connect facts, but this may also lead to longer stories and introduce subjectivity.}

\rvm{\textbf{Story customization:} 
\toolName{} currently supports only coarse--grained user control through the provided filters, such as the number of clusters.
Beyond these controls, 
the system does not yet support deeper narrative shaping, 
such as aggregating additional facts or expanding contextual information, to match different sensemaking intents.
Future work could move toward more dynamic and user-steerable storytelling approaches, for example, systems like Socrates~\cite{guande2024socrates} that adapt data stories based on user feedback, enabling stories to be iteratively refined, such as by adjusting the amount of contextual scaffolding or narrative detail.}

\rvm{\textbf{Sparse, missing, and heterogeneous data:}
When only a few quantitative data points are available, narrative visualizations might become less expressive and inadvertently mislead interpretation (\eg missing intermediate values masking volatility).
Moreover, the extracted quantities may be heterogeneous across sources (\eg following different population definitions), which could reduce comparability even if each value is accurate. 
Future work could improve merging and normalization rules (\eg aligning definitions where possible), 
explicitly surface missing data through visual cues, or explore approaches that represent absent values and uncertainty (\eg as in ChartifyText~\cite{zhang2024chartifytext}). 
However, richer alignment and uncertainty handling may increase computational overhead and generation latency.}

\section{Conclusion}

We present \toolName{}, an automated system that retrieves and analyzes a collection of online articles in response to a user query and delivers a coherent and meaningful visual data story tailored to the user’s informational needs through scrollytelling. 
Our framework comprises two main modules: 
the Data Fact Extraction and Organization module and the Visual Storytelling module, each designed to address the unique challenges of working with unstructured textual data. 
We evaluate \toolName{} through a quantitative analysis and  two user studies, 
demonstrating that \toolName{} effectively extracts accurate facts, organizes them into meaningful narratives, and presents them in a visually compelling and user-friendly manner.
In the future, we plan to enhance system usability by reducing latency through more efficient, high-performing LLMs and by integrating fact-checking techniques to increase the trustworthiness.
We also believe this work opens new directions for search interfaces, where narrative storytelling can transform user engagement with search results by incorporating not only objective facts but also subjective opinions.

\section*{Acknowledgments}
This project is supported by NTU Start Up Grant awarded to Yong Wang.

\bibliographystyle{IEEEtran}
\bibliography{main}

% \vspace{-3em}
\begin{IEEEbiography}[{\includegraphics[width=1in,height=1.2in,clip,keepaspectratio]{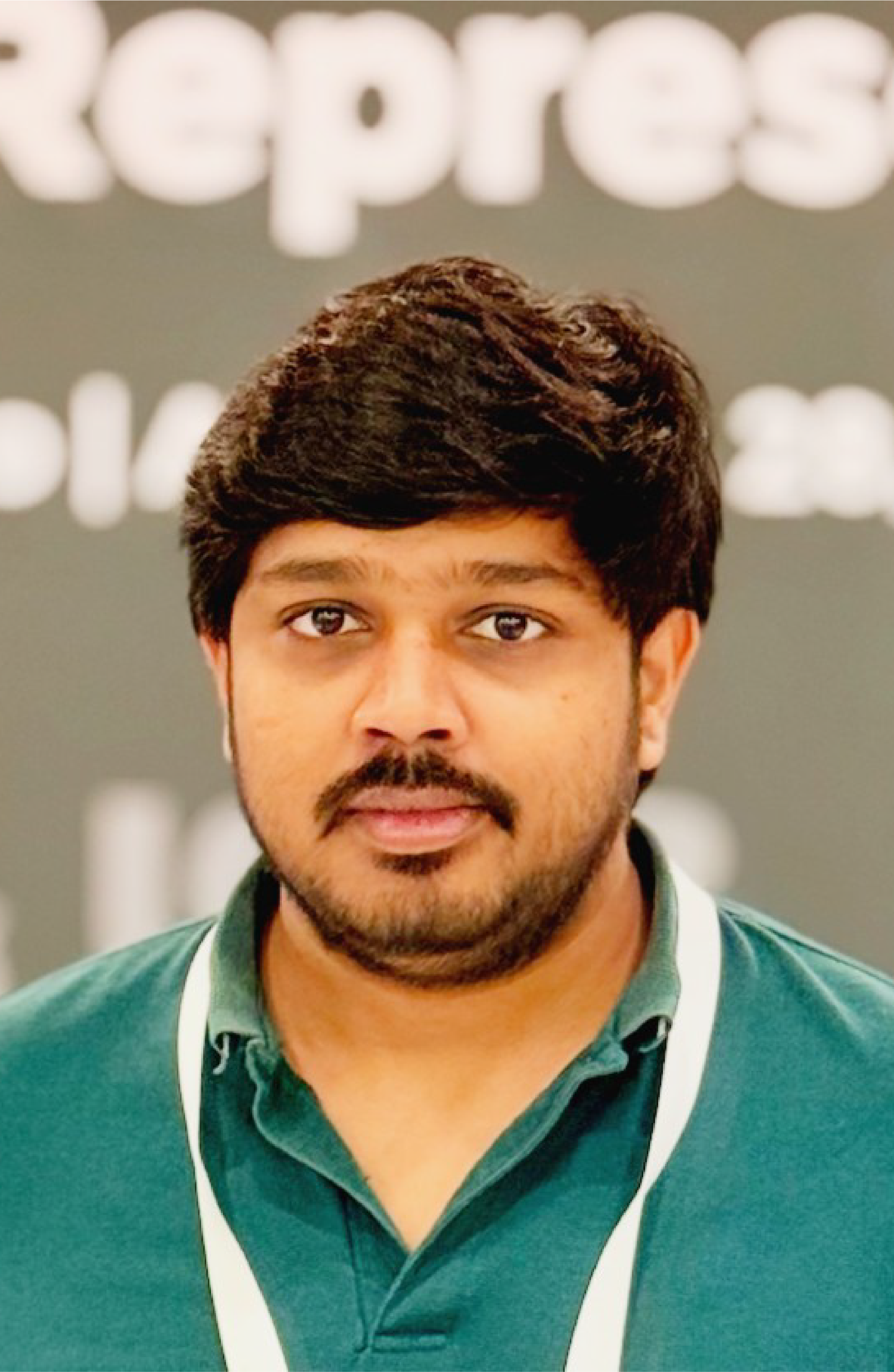}}]{Manusha Karunathilaka} is currently a Ph.D. candidate in School of Computing and Information Systems at Singapore Management University (SMU). His research interests include data visualization, storytelling, and human–computer interaction, with a particular focus on automated visual story generation. He received his bachelor’s degree in Computer Science and Engineering from the University of Moratuwa, Sri Lanka. For more details, kindly visit \url{https://manusha-karunathilaka.com}.
\end{IEEEbiography}
 
% \vspace{-3em}
\begin{IEEEbiography}[{\includegraphics[width=0.8in,height=1.2in,clip,keepaspectratio]{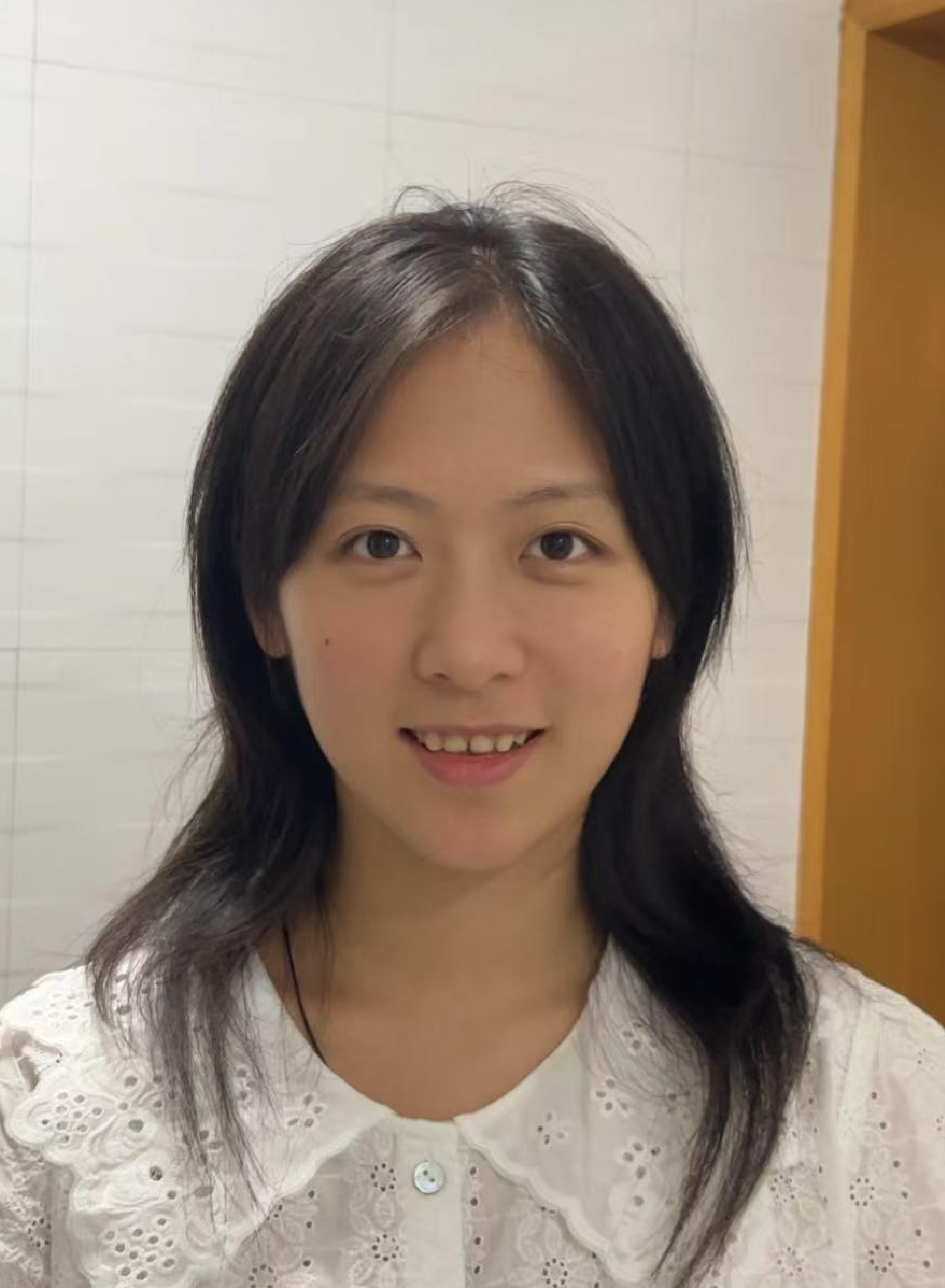}}]{Litian Lei} is currently a master student in Artificial Intelligence at the College of Computing and Data Science, Nanyang Technological University. Her research interests include data visualization, human-computer interaction and computer vision. She received her bachelor’s degree in Information and Communications Technology from Royal Institute of Technology in Sweden.
\end{IEEEbiography}
 
% \vspace{-3em}
\begin{IEEEbiography}[{\includegraphics[width=0.8in,height=1.2in,clip,keepaspectratio]{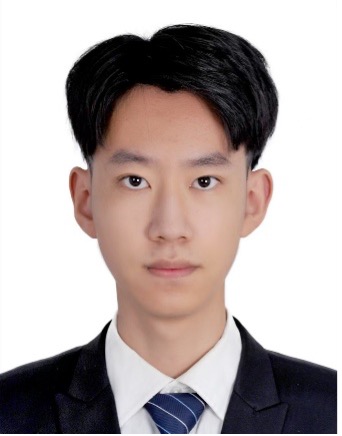}}]{Yiming Gao} is currently an undergraduate student at the College of Computing and Data Science, Nanyang Technological University. His research interests lie in the areas of Large Language Model (LLM) evaluation pipelines and reinforcement learning (RL). 
\end{IEEEbiography}

% \vspace{-3em}
\begin{IEEEbiography}[{\includegraphics[width=1in,height=1.2in,clip,keepaspectratio]{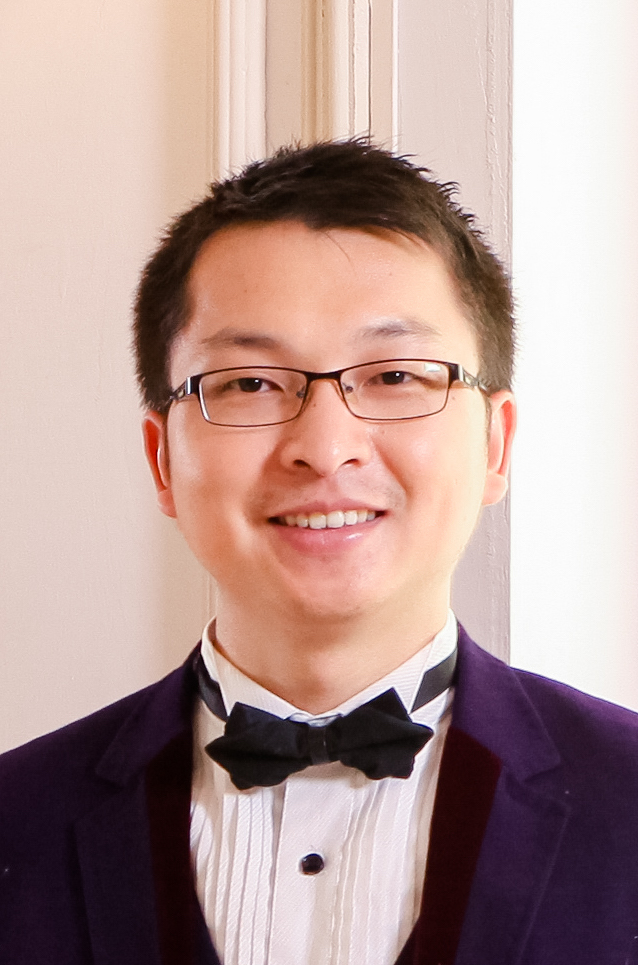}}]{Yong Wang} is currently an assistant professor in the College of Computing and Data Science, Nanyang Technological University. Before that, he worked as an assistant professor at Singapore Management University from 2020 to 2024. His research interests include data visualization, HCI and human-AI collaboration, with an emphasis on their application to FinTech, quantum computing and online learning. He obtained his Ph.D. in Computer Science from Hong Kong University of Science and Technology. He received his B.E. and M.E. from Harbin Institute of Technology and Huazhong University of Science and Technology, respectively. For more details, please refer to \url{http://yong-wang.org}.
\end{IEEEbiography}

% \vspace{-2.5em}
\begin{IEEEbiography}[{\includegraphics[width=0.8in,height=1.2in,clip,keepaspectratio]{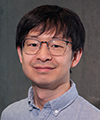}}]{LI Jiannan} is currently an assistant professor in School of Computing and Information Systems at Singapore Management University. 
His research interests include Human-Computer Interaction and Human-Robot Interaction, with an emphasis on improving interactions between humans and intelligent agents.
He obtained his Ph.D. in Computer Science from University of Toronto. 
He received his MSc in Computer Science from University of Calgary and BEng in Automation and Control from Southeast University, China.
For more details, please refer to \url{https://jchrisli.github.io/}.
\end{IEEEbiography}

\vfill

\end{document}